\documentclass[12pt]{article}
\pdfoutput=1
\usepackage{putex}

\usepackage{feyn}
\usepackage[vcentermath]{youngtab}
\usepackage{subfig}
\usepackage{lscape}
\usepackage{braket}
\usepackage{comment}
\usepackage{graphicx}
\usepackage{epstopdf}
\usepackage{enumerate}
\usepackage{cite}
\usepackage{tensor}
\usepackage{slashed}
\usepackage{amsmath}
\usepackage{amssymb}
\usepackage{mathrsfs}
\usepackage{lgrind}

\usepackage{bbm}
\usepackage{hyperref}

\numberwithin{equation}{section}

\newcommand{\detq}{\det\nolimits_q}

\newcommand {\be} {\begin {equation}}
\newcommand {\ee} {\end {equation}}

\newcommand {\bes} {\begin {equation*}}
\newcommand {\ees} {\end {equation*}}

\def\CH{{\cal H}}

\newcommand{\beq}{\begin{equation}}
\newcommand{\eeq}{\end{equation}}

\newcommand{\sq}{SL_q(2)}
\newcommand{\mO}{\mathcal{O}}

\newcommand{\cop}{{\bf coproduct}}
\newcommand{\anti}{{\bf antipode}}
\newcommand{\cou}{{\bf co-unit}}
\newcommand{\osq}{\mO(\sq)}
\newcommand{\uq}{U_q(\mathfrak{sl}_2)}
\newcommand{\lan}{\langle}
\newcommand{\ran}{\rangle}
\newcommand{\su}{SU_q(1,1)}
\newcommand{\us}{U_q(\mathfrak{su}_{1,1})}
\newcommand{\osu}{\mO(\su)}

\def\be{ \begin{equation} }
\def\ee{ \end{equation} }

\usepackage{authblk}
\begin{document}

\institution{NYU}{Center for Cosmology and Particle Physics, New York University, New York, NY 10003, USA}

\institution{NYUAD}{New York University Abu Dhabi, P.O. Box 129188, Abu Dhabi, United Arab Emirates}

\title{Holography on the Quantum Disk}

\authors{Ahmed Almheiri\worksat{\NYU}\worksat{\NYUAD}, Fedor K. Popov\worksat{\NYU}}

\abstract{Motivated by recent study of DSSYK and the non-commutative nature of its bulk dual, we review and analyze an example of a non-commutative spacetime known as the quantum disk proposed by L. Vaksman. The quantum disk is defined as the space whose isometries are generated by the quantum algebra $\us$. We review how this algebra is defined and its associated group $\su$ that it generates, highlighting its non-trivial coproduct that sources  bulk non-commutativity. We analyze the structure of holography on the quantum disk and study the imprint of non-commutativity on the putative boundary dual.}

\date{}
\maketitle

\tableofcontents

\section{Introduction}

Everything in quantum gravity should be quantum, including spacetime. A notion of a quantum spacetime is described by a non-commutative geometry i.e. one whose coordinates do not commute with one another. Examples of such spacetimes have appeared in several areas in theoretical physics including string theory\cite{Seiberg:1999vs}, matrix models \cite{Connes:1997cr} and quantum field theories on non-commutative spaces \cite{Gross:2000ph,Gross:2000ss,Gross:2000wc,Alvarez-Gaume:2003lup}.

Recently, non-commutative geometry has showed up in the bulk picture of double scaled SYK model \cite{Berkooz:2018jqr, Berkooz:2018qkz, Berkooz:2022mfk, Lin:2023trc}. The double scaling limit is the limit where the number of fermions $N$ and the order of the fermion interaction in the SYK Hamiltonian $p$ are taken to infinity while keeping $\lambda \equiv 2p^2/N$ fixed \cite{Cotler:2016fpe}. The bulk picture that emerges is described in terms of chords anchored to the boundary \cite{Berkooz:2018jqr,Berkooz:2018qkz,Milekhin:2023bjv}. These chords provide an unusual, primitive version of spacetime where distances in the bulk are measured in units of $\lambda$, the effective Planck length in the bulk (or ratio between the Planck and AdS scales). Smooth spacetime emerges in the small $\lambda$ limit in which the number of chords proliferate and smooth spacetime emerges.

At finite temperature of DSSYK, symmetry algebra of the spacetime is a subalgebra of $\us$ \cite{Lin:2023trc}. The “quantum” aspect of the algebra is the presence of a non-trivial coproduct. The action of an infinitesimal generator on elements of tensor product of $\us$ modules does not satisfy the Leibniz rule, but instead has a ``tailed'' action given by
\begin{align}
E[f \otimes g] = E[f] \otimes g + K[f] \otimes E[g],
\end{align}
where $E,K \in \us$. The discreteness of the chord spacetime is intimately related to this coproduct structure.

Surprisingly, this non-trivial coproduct has a remnant in the $\lambda \rightarrow 0$ limit at finite temperature in the regime of large $p$ limit of SYK. As shown in \cite{Lin:2023trc}, the two point function in this regime is not conformal, yet it is related to a conformal correlator through a non-symmetric coordinate transformation on the coordinates of the two points:
\begin{align}
    \cos^{-2 \Delta}\left[{\pi v \over 2}\left(1 - {\theta_+ - \theta_- \over \pi}\right)\right] = \sin^{-2 \Delta} \left[ {\varphi_+ - \varphi_- \over 2}\right],
\end{align}
where $\varphi_\pm =  \pm {\pi \over 2} + v \left( \theta_\pm \mp {\pi \over 2} \right)$. The coordinates $\varphi_\pm$ are the coordinates on the ``fake disk'' whose boundary length is $\beta /v$. The fake disk coordinates makes the $SU(1,1)$ symmetry manifest. For instance, the correlators invariant under the action of $E[\mO] \equiv \cos \varphi \partial_\varphi \mO - \Delta \sin \varphi \mO$. If we express this symmetry in terms of physical disk coordinates $\theta_\pm$, the generator $E$ will have to act differently on the $\mO(\theta_+)$ and $\mO(\theta_-)$ due of the asymmetry in the transformation between $\varphi_\pm$ and $\theta_\pm$. However, this asymmetry can be compensated by assigning a non-trivial coproduct to the action of $E$ by introducing another operator $P$ whose actions in terms of  physical coordinates is
\begin{align}
    &E[\mO(\theta_+) \mO(\theta_-)] = E[\mO(\theta_+)] \mO(\theta_-) + P[\mO(\theta_+)] P \cdot E[\mO(\theta_-)],
\end{align}
where
\begin{align}
    &E[\mO(\theta)] \equiv v^{-1} \cos\left[{\pi \over 2}  + v \left( \theta - {\pi \over 2} \right)\right] \partial_\theta \mO(\theta) - \Delta \sin\left[{\pi \over 2}  + v \left( \theta - {\pi \over 2} \right)\right] \mO(\theta), \\
    &P[\mO(\theta)] \equiv \mO(\theta - {\pi(1-v)/ v}).
\end{align}

These observations motivate the study of standard examples involving a non-trivial coproduct, hence this paper on the quantum disk and groups. This paper will analyze the structure of holography on a non-commutative version of the hyperbolic disk, also known as the quantum disk that was studied extensively by L. Vaksman \cite{2008arXiv0803.3769V,shklyarov1998function1, shklyarov1999function2, shklyarov1998function3, shklyarov1998function4, shklyarov1999function5}. This is a non-commutative spacetime whose symmetry group is $SU_q(1,1)$.

This topic is sufficiently foreign to our  community that we felt that we should dedicate a large fraction of this paper to review L. Vaksman's seminal works on the quantum disk. Hence, sections 2 is a pedagogical review of \cite{2008arXiv0803.3769V,shklyarov1998function1, shklyarov1999function2, shklyarov1998function3, shklyarov1998function4, shklyarov1999function5} describing how to define the quantum disk and its symmetry quantum group. In section 3, we describe the mechanics of holography, extracting boundary correlation functions and induced quantum symmetry transformations of boundary primary operators. In section 4, we study symmetry aspects of the putative boundary dual of the quantum disk. We find the three point function using symmetry and match it to a bulk Witten diagram.

\section{A review of the quantum  disk}

In this section we describe how to define the quantum disk \cite{vaksman2010quantum,shklyarov1998function1, shklyarov1999function2, shklyarov1998function3, shklyarov1998function4, shklyarov1999function5}, a $q$-deformed version of the hyperbolic disk, as the space whose symmetry group is the quantum group $\su$ \cite{klimyk2012quantum,kassel2012quantum,woronowicz1998compact,manin1996quantum}.  The coordinates on the quantum disk will be noncommutative, a property that descends from the noncommutativity of the matrix elements of group elements of $\su$. We will not describe the intrinsic geometric structure of the quantum disk, which is obscured by its noncommutative nature, but rather focus on the properties of functions of its coordinates.

\subsection{$\su$: the symmetry group of the quantum disk}

To understand the symmetry group of the quantum disk $\su$, we start with the quantum deformation of the $SL(2,\mathbb{C})$, called $\sq$. Elements of $\sq$ are all $2\times 2$ matrices, whose entries satisfy the following relations
\begin{align}
    \sq = \left\{ \begin{pmatrix}
a & b\\
c & d 
\end{pmatrix} \Bigg| \ \  \   \begin{matrix}
ab = q ba, \ \ \  bd = qdb,  \ \ \  ad = da + (q-q^{-1}) cb\\
ac = q ca, \ \ \  cd = q dc,  \ \ \ bc = cb,\  \ \ ad - qbc = 1
\end{matrix} \right\}, \label{slq}
\end{align}
The combination $\mathrm{det}_q \equiv ad - q bc$ is called the quantum determinant and belongs to the center of the $a,b,c,d$ algebra. Without the condition $\mathrm{det}_q = 1$, \eqref{slq} defines $GL_q(2)$. The above relations generate an ideal and are determined by a solution to an $R$-matrix equation \eqref{eq:yba}. We describe this in details in appendix \ref{rmatrix}.

The equation \eqref{slq} is the defining representaion of $\sq$ \cite{faddeev1989quantum}.  It is sometimes called its ``coordinate representation" since the matrix elements define coordinates parameterizing group elements on the group manifold. Just like normal groups, this group includes the identity element $e$ where $a = d = 1, b = c = 0$, which  satisfies the relations \eqref{slq}. Thus, one can define a map $\epsilon$ that sets any group element $g \in \sq$ to the identity. It is known as the \cou \ $\epsilon(g) = e$. 

We can define the inverse of the above elements as
\begin{align}
    g  = \begin{pmatrix}
a & b\\
c & d 
\end{pmatrix}
\rightarrow g^{-1}  = \begin{pmatrix}
d & -q^{-1}b\\
-q c & a 
\end{pmatrix}.
\end{align}
Formally, we have $g\cdot g^{-1}=g^{-1}\cdot g = e$. The map from a group element to its inverse is known as the \anti\ $S:g \rightarrow g^{-1}$. Formally, on the entries of matrix defining the group element we have
\begin{align}
    S(a) = d, \ S(b) = - q^{-1}b,\ S(c) = -q c, \ S(d) = a \label{antip}.
\end{align}

Group multiplication can also be defined for $\sq$. It is a map from the matrix elements of a pair of group elements $g_1$ and $g_2$ to their product $g_1 g_2$ in a way that preserves the ideal \eqref{slq}. To define this, first we should think of the matrix elements $a,b,c,d$  as functions on the group manifold that assign a value (abstractly) to each group element, i.e. the matrix elements of a group element $g$ are $a(g), b(g), c(g), d(g)$. Then, the multiplication can be defined as a map $\hat{\times}$ that takes matrix elements of two group elements to matrix  elements of a product of these elements 
\begin{gather}
\begin{pmatrix}
a(g_1) & b(g_1)\\
c(g_1) & d(g_1) 
\end{pmatrix} \hat{\times} \begin{pmatrix}
a(g_2) & b(g_2)\\
c(g_2) & d(g_2) 
\end{pmatrix} \\ = 
\begin{pmatrix}
a(g_1)\otimes a(g_2) + b(g_1) \otimes c(g_2) & a(g_1)\otimes b(g_2) + b(g_1) \otimes d(g_2)\\
c(g_1)\otimes a(g_2) + d(g_1) \otimes c(g_2) & c(g_1)\otimes b(g_2) + d(g_1) \otimes d(g_2) 
\end{pmatrix}
\equiv \begin{pmatrix}
a(g_{1}g_{2}) & b(g_{1}g_{2})\\
c(g_{1}g_{2}) & d(g_{1}g_{2}) 
\end{pmatrix}.\notag
\end{gather}
Thus we have mapped the matrix element functions $a,b,c,d$ to the functions acting on two copies of the group manifold.
This operation of mapping functions on one copy of a  space to the functions on two copies is called the \cop, and is denoted by $\Delta(\cdot)$. For $\sq$ we have
\begin{align}
    \Delta(a) = a \otimes a + b \otimes c, \\
    \Delta(b) = a \otimes b + b \otimes d, \\
    \Delta(c) = c \otimes a + d \otimes c, \\
    \Delta(d) = c \otimes b + d \otimes d.
\end{align}
The \cop\ is an algebra homomorphism.

One is often interested in defining functions on the group manifold. This space of functions is a vector space spanned by monomials $a^i b^j c^k d^l$ with integer powers and is denoted as $\mO(\sq)$. The notions of {\bf co-unit, antipode} and {\bf coproduct}  extend to this algebra in the obvious way. The \cou \ implements $\epsilon: \osq \rightarrow \mathbb{C}$, while the {\bf \cop}\ implements $\Delta: \mO(\sq) \rightarrow \mO(\sq) \otimes \mO(\sq)$. We can introduce ``reversed" maps that are compatible with these maps. Thus, we define as the {\bf unit} that maps a constant to the identity element of the algebra $\eta: \mathbb{C} \rightarrow \osq$ (it also maps a constant to the identity  of the group $\sq$) and a multiplication called the {\bf product} that maps a tensor product to a single copy $m: \osq \otimes \osq \rightarrow \osq$ acting as as $m(a \otimes b) = ab$.  These structures together make $\osq$ a {\bf Hopf algebra}.

To get $\su$ we need to impose reality conditions on $\sq$. For $q = 1$, both are Mobius transformations on the complex plane, but $SU(1,1)$ preserves the unit disk. Defining complex conjugation makes $\osq$ into a { \bf Hopf $*$-algebra} with an involution $*$ that satisfies $(ab)^* \rightarrow b^* a^*$, i.e. it's an anti-involution. We choose the reality condition for $\su$ to be $g^* = \eta S(g) \eta$ where $\eta = \diag\{1,-1\}$, giving
\begin{align}
    \begin{pmatrix}
a^* & c^*\\
b^* & d^* 
\end{pmatrix}
=\begin{pmatrix}
d & q^{-1} b\\
q c & a 
\end{pmatrix}, \label{conjgroup}
\end{align}
This defines a conjugation on all elements in $\osu$.

\subsection{$\us$:  generators of  $\su$}

One can define a set of generators on the group manifold of $\su$ that formally implements an infinitesimal translation. These generators are usually labelled $K,E,F$ and they satisfy an algebra known as $\us$ (more precisely, this is the vector space spanned by polynomials of $K,E,F$). Their action on the elements of $SU_q(1,1)$ is defined through the pairing 
\begin{align}
   q^{-1} \langle K,a \rangle =  q \langle K,d\rangle = q^{-1/2}\langle E, c \rangle = q^{1/2} \langle F, b \rangle = 1\,, \label{eq:pair}
\end{align}
Hence, on a group element we have
\begin{align}
    \lan K, g \ran =  \begin{pmatrix}
q & 0\\
0 & q^{-1} 
\end{pmatrix}, \quad 
 \lan E, g \ran =  \begin{pmatrix}
0 & 0\\
q^{1/2} & 0 
\end{pmatrix}, \quad 
 \lan F, g \ran =  \begin{pmatrix}
0 & q^{-1/2}\\
0 & 0 
\end{pmatrix}. \label{gen}
\end{align}
 The pairing extends to relate all elements of $\us$ to $\mO(\su)$, and establishes a duality between the two algebras. To do so, it must satisfy a set of constraints (see appendix \eqref{sec:appa}).

One constraint is it must respect the algebra relations (or {\bf ideal}) in \eqref{slq} once the pairing is extended to all elements of $\osu$. For instance, we must have $ \lan E, ac \ran = q \lan E, ca \ran$. There are multiple ways of extending the pairing, one of which is to impose that 
\begin{align}
   \lan K, A B \ran &= \lan K,A \ran \lan K,B \ran, \label{k}\\
   \lan E, A B \ran &= \lan E,A \ran \lan 1,B \ran + \lan K,A \ran \lan E,B \ran, \label{e} \\
   \lan F, A B \ran &= \lan F,A \ran \lan K^{-1},B \ran + \lan 1,A \ran \lan F,B \ran, \label{f}
\end{align}
for any $A,B \in \osq$. The pairing with the identity implements  the \cou, $\lan 1,B \ran = \epsilon(B)$. These relations define a \cop\ for $\us$ given by 
\begin{align}
    \Delta(K) &= K\otimes K, \\
    \Delta(E) &= E\otimes 1 + K \otimes E,\\
    \Delta(F) &= F\otimes K^{-1} + 1\otimes F,
\end{align}
provided that the pairing satisfies
\begin{align}
    \lan X \otimes Y, A \otimes B \ran = \lan X, A \ran \lan Y, B \ran,
\end{align}
where $X,Y \in \us$. As a result, the pairing can be shown to satisfy
\begin{align}
    \lan \Delta(XY), A \otimes B \ran = \lan X \otimes Y, \Delta(AB) \ran.
\end{align}

The generators $K,E,F$ must satisfy a set of reality conditions compatible with  the reality condition of $\osu$. This can be defined using that the pairing satisfies
\begin{align}
    \lan X^*,A \ran = \lan X, S(A)^* \ran^*.
\end{align}
Using the $\cop$ from $K,E,F$, the {\bf antipode} \eqref{antip} and the relations \eqref{conjgroup}, implies the following reality conditions
\begin{align}
    K^* = K, \ E^* =  -K F, \ F^* =  -E K^{-1}.
\end{align}
Further conditions on the pairing can be used to define a {\bf unit, co-unit, antipode, a product} for the algebra of $K,E,F$ that makes it a Hopf $*$-algebra. We review these additional structures in appendix \ref{hopf}. So far, we have established a duality between two Hopf $*$-algebras $\mO(SU_q(1,1))$ and freely generated algebra of $K,E,F$. 

Finally, the pairing \eqref{eq:pair} is not faithful and has a kernel generated by the relations, or {\bf ideal}, between $K,E,F$,
\begin{align}
    KE = q^2 EK , \ \ KF = q^{-2} FK, \ \   EF - FE = {K - K^{-1}\over q - q^{-1}}.
\end{align}
See  appendix \ref{idderiv} for a derivation of this kernel. Imposing these on the algebra of $K,E,F$ defines the algebra $\us$. From this one can deduce a Casimir that commutes with the generators
\begin{align}
    C_q = EF + {q^{-1}(K-1) + q (K^{-1}-1) \over (q^{-1} - q)^2}. \label{casimir}
\end{align}
The action of the Casimir on tensor product representations descends from the \cop\ of the generators, and is given by
\begin{gather}
    \Delta(C_q)  = 
    C_q \otimes K^{-1} + K \otimes C_q + E \otimes F + KF \otimes EK^{-1} - \frac{(q^{-1}+q)(K-1)\otimes (K^{-1}-1)}{(q-q^{-1})^2}. \label{casimircop}
\end{gather}

\subsection{Coordinates on the quantum disk and their algebra}

Having defined the symmetry group of the quantum disk and its associated generators, we turn to defining the quantum disk itself. Recall that the standard hyperbolic disk can be obtained from a quotient of the $SL_2(\mathbb{R})$ group manifold, see \cite{kitaev2018notes}. Similarly, as discussed in \cite{shklyarov1998function3}, the quantum disk can be defined as the quotient of $\su$ by the right action of $K$\footnote{For the sake of representation we introduce another copy of $\su$ as a homogeneous space of $\su$ with coordinates $t_{ij}$.},
\begin{gather}
    \su = \left\{ \begin{pmatrix}
t_{11} & t_{12}\\
t_{21} & t_{22} 
\end{pmatrix} \Bigg|   \begin{matrix}
t_{11} t_{12} = q t_{12} t_{11}, \  t_{12}t_{22} = qt_{22}t_{12},\  t_{11}t_{21} = q t_{21}t_{11}, \ t_{21}t_{22} = q t_{22}t_{21},  \\   t_{12}t_{21} = t_{21} t_{12},\  t_{11}t_{22} = t_{22}t_{11} + (q-q^{-1}) t_{21} t_{12}, \ t_{11} t_{22} - q t_{12}t_{21} = 1, \
 \  
\end{matrix} \right\} \notag \\
\mathbb{D}_q = \su \slash K. \label{coordsl2}
\end{gather}
The action (or coaction) of $\su$ on the disk correspond to left matrix multiplication $t' = g \times t$ that maps to the tensor product as
\begin{align}
    t_{11}  \rightarrow t'_{11} = a \otimes t_{11} + b \otimes t_{21}, \\
    t_{12}  \rightarrow t'_{12} = a \otimes t_{12} + b \otimes t_{22}, \\
    t_{21}  \rightarrow t'_{21} = c \otimes t_{11} + d \otimes t_{21}, \\
    t_{22}  \rightarrow t'_{22} = c \otimes t_{12} + d \otimes t_{22}.
\end{align}
The elements $t_{ij}'$ have the same algebra as $t_{ij}$. It will be useful later to note that $\detq(t') = \detq(g\otimes t) = \detq g \otimes \detq t$ where $\detq t= t_{11}t_{22}-q t_{12}t_{21}$.

It is useful to analyze the action of generators on $t_{ij}$. The action of the generators on $t_{ij}$ is obtained using \eqref{gen} implying the following transformation rules
\begin{align}
    &K(t_{11})  = q t_{11}, \ K(t_{12}) = q t_{12}, \ K(t_{21}) = q^{-1} t_{21}, \ K(t_{22}) = q^{-1} t_{22}, \\
    &E(t_{11})  = 0, \ E(t_{12}) = 0, \ E(t_{21}) = q^{1/2}  t_{11}, \ E(t_{22}) = q^{1/2}t_{12}, \\
    &F(t_{11})  = q^{-1/2}t_{21}, \ F(t_{12}) = q^{-1/2} t_{22}, \ F(t_{21}) = 0, \ F(t_{22}) = 0.
\end{align}
To obtain the action of $\uq$ on arbitrary products and powers of $t_{ij}$, we follow the same rules as those defined in the pairing in the previous section. The {\bf coproduct} then implements the action on products of $t_{ij}$.

\subsubsection{Disk coordinates}

Now we can implement the quotient to define the disk coordinates $z = q^{-1} t^{-1}_{21}t_{11}, z^* = t_{22} t_{12}^{-1}$, which are indeed invariant under the rescaling of $t_{ij}$ by the action of $K$. Using the $SU_q(1,1)$ reality conditions $t_{11}^* = t_{22}, t_{12}^* = q t_{21}$ we see that $z,z^*$ are conjugate to each other. This also follows from the conjugation relations \eqref{conjgroup} and by conjugating either side of the transformations $t_{ij} \rightarrow t_{ij}'$. Under the above transformation rules, we have 
\begin{align}
    z \rightarrow w =  q^{-1} t'^{-1}_{21}t'_{11} =&  (c \otimes t_{11} + d \otimes t_{21} )^{-1}(a \otimes t_{11} +  b\otimes t_{21} )\notag\\
    =& q^{-1} ( q c z + d)^{-1} ( qaz + b),
\end{align}
which is a usual Mobius transformation. The tensor product symbol is dropped from the last equation for brevity. As a check, one can show that this mapping preserves the boundary; suppose that $w,w^*$ are constructed from the matrix $t'$ just as $z,z^*$ are constructed from $t$. Boundary points in $z$ correspond to $zz^* = 1$ which satisfy $\detq t = 0$ thus implying that $\detq t' = 0$ and  $ww^* = 1$\footnote{Since the disk, classical or quantum, is strictly an open set, it doesn't include the boundary and hence $\detq$ is allowed to vanish.}.

The action of the generators on $z,z^*$ can be deduced from their action on $t_{ij}$ and using the \cop, which gives
\begin{align}
    &K(z) = q^2 z, \ E(z) = - q^{5/2} z^2, \ F(z) = q^{-3/2}, \label{efkaction1}\\
    &K(z^*) = q^{-2} z^*, \ E(z^*) = q^{1/2}, \ F(z^*) = -q^{1/2}z^{*2}.\label{efkaction2}
\end{align}
Again, using the \cop \ we can extend this action to powers of $z,z^*$. For ``(anti)holomorphic" functions we have
\begin{align}
    &K(g(z)) = g(q^2 z), \ E(g(z)) = - q^{5/2} z^2 {g(z) - g(q^2 z) \over z - q^2z}, \ F(g(z)) = q^{-3/2}{g(z) - g(q^{-2}z) \over z - q^{-2}z},\label{efkaction3} \\
    &K(g(z^*)) = g(q^{-2} z^*), \ E(g(z^*)) = q^{1/2}  {g(z^*) - g(q^{-2} z^*) \over z^* - q^2z^*}, \ F(g(z)) = -q^{1/2}z^{* 2}{g(z^*) - g(q^{2}z^*) \over z^* - q^{2}z^*}. \label{efkaction4}
 \end{align}

The quantum disk has the interesting property of being noncommutative, namely that $z,z^*$ do not commute, which descends from the noncommutativity of the matrix elements \eqref{coordsl2}. Starting from $t_{11}t_{22} = t_{22}t_{11} + (q-q^{-1}) t_{21} t_{12}$, we can multiply from the right(left) with $t_{21}^{-1}(t_{12}^{-1})$ respectively, and then use the remaining relations in \eqref{coordsl2} to commute things around until we have an expression  involving only $z,z^*$. The resulting relation is
\begin{align}
    z^* z = q^2 z z^* + 1 - q^2.
\end{align}
This relation can also be obtained using $R$-matrix methods. As discussed in appendix \ref{rmatrix}, the universal $R$-matrix for $\us$ is given by
\begin{align}
    \breve{R}(\cdot) = \exp_{q^2}\left((q^{-1} - q)F\otimes E \right)  q^{-{H \otimes H \over 2}} (\cdot), \label{rm}
\end{align}
which in particular gives
\begin{align}
    z^* z = m \breve{R}(z \otimes z^*) = q^2 z z^* + 1 - q^2,
\end{align}
where the $m:\mathbb{C}_q[z,z^*] \otimes \mathbb{C}_q[z,z^*] \to \mathbb{C}_q[z,z^*]$ multiplies the elements of the tensor product together\footnote{We define the algebra structure on the tensor products of two algebras $A_{1,2}$ as $(a_1 \otimes b_1) (a_2 \otimes b_2) = a_1 a_2 \otimes b_1 b_2$.}. 

The algebra of $z,z^*$ is denoted as $\mathbb{C}_q[z,z^*]$. An important element of this algebra is an element $y = 1 - z z^*$ that satisfy the following commutation relations
\begin{gather}
    z y = q^{-2} y z, \quad z^* y = q^{2} y z^*.
\end{gather}
It is useful to show connection of $\mathbb{C}_q[z,z^*]$ with the $q$-deformed harmonic oscillator. Thus, we notice that 
\begin{gather}
    \hat{a} = \frac{z}{\sqrt{1-q^2}}, \quad \hat{a}^\dagger = \frac{z^*}{\sqrt{1-q^2}}, \quad
     \hat{a}^\dagger \hat{a} - q^2 \hat{a} \hat{a}^\dagger = 1,
\end{gather}
thus any element of $f \in \mathbb{C}_q[z,z^*]$ could be thought as an operator acting in the Hilbert space of $q$-deformed harmonic oscillator.  That Hilbert space is spanned by vectors $\ket{n}, n \in \mathbb{N}_{\geq 0}$ and the action is given by
\begin{gather}
    z \ket{n} = \ket{n+1} , \quad z^* \ket{n} = \left(1-q^{2n}\right)\ket{n-1}, \quad y \ket{n} = q^{2n}\ket{n}, \label{eq:qharmrep}
\end{gather}
one can show that this mapping is actually an isomorphism and any function $\psi(y)$ of $y$ could be given just by assigning the values of $\psi(q^{2n})$. In some loose sense, the quantum disk is a set of concentric circles at discrete radii of constant $y$ that accumulate up to the unit circle.

\subsubsection{Differential $q$-calculus}

Various notions of differential calculus can be extended to the non-commutative setting. One can define a notion of a differential operator $d$  that satisfies the usual Leibnez rule $dz^2 = z dz + dz\,  z$. The differential form $dz$  satisfies $d(dz) =0$, which follows from $d^2=0$. The action on $z^*$ follows similarly.

The action of the generators of $\us$ on the differentials follows from the Leibnez rule along the commutations above to give
\begin{align}
     &K(dz) = q^2 dz, \ E(dz) = - q^{5/2} dz z (1 + q^{-2}), \ F(dz) = 0,\\
    &K(dz^*) = q^{-2} dz^*, \ E(dz^*) = 0, \ F(dz^*) = -q^{1/2}dz^* z^{*}(1 + q^2).
\end{align}
The algebra between $dz,dz^*, z,z^*$ follows from the $R$-matrix in \eqref{rm}. We have
\begin{align}
    &zdz = q^{-2} dz \, z, \ z^* dz^* = q^{2} dz^* \, z^*, \\ 
    &zdz^* = q^{-2} dz^* z, \ z^* dz = q^2 dz \, z^*. \label{dzr}
\end{align}
The differentials allow us to define differential operators acting on any function $h(z,z^*)$ on the quantum disk through
\begin{align}
    dh(z,z^*) &= dz {1 \over q^{-2}z - z} \left( {h(q^{-2}z,z^*) - h(z,z^* ) } \right) + dz^* \left( h(q^{-2}z,z^*) - h(q^{-2}z,q^{2}z^* ) \right) { 1 \over z^* - q^2z^*}\notag \\
    &\equiv dz {\partial  h\over \partial z}  + dz^* {\partial  h \over \partial z^*}\equiv \partial h + \bar{\partial} h. 
\end{align}
This defines the derivative operators ${\partial \over \partial z}, {\partial \over \partial z^*}$, and holomorphic and anti-holomorphic differentials  $\partial h, \bar{\partial} h$. % = dz {\partial h \over \partial z}, \, \bar{\partial} h = dz^* {\partial h \over \partial z^*}$ 
This means we can write $d = \partial + \bar{\partial}$. It's not hard to check that these derivatives satisfy the following algebra
\begin{align}
    & \ \ \   {\partial \over \partial z} z^* =  q^{2}z^*{\partial \over \partial z}, \ {\partial \over \partial z^*} z =  q^{-2}z{\partial \over \partial z^*}, \\
    &{\partial \over \partial z} z = 1 + q^{-2} z {\partial \over \partial z}, \ {\partial \over \partial z^*} z^* = 1 + q^{2} z^* {\partial \over \partial z^*}.
\end{align}
The Casimir \eqref{casimir} can be expressed in terms of these derivatives%. Using the form the Casimir \eqref{casimir} and its \cop \eqref{casimircop}, its given by
\begin{align}
    C_q = q^{-1} \left(1-z z^*\right)^2 \frac{\partial}{\partial z} \frac{\partial}{\partial z^*}. \label{casimirdiff}
\end{align}
We check in the appendix \eqref{app:casandder} that the two sides of this equation have the same action on any monomial $z^n z^{*m}$.

Next we define how to perform integrals over the entire quantum disk. Such an integral can be thought of as a map of functions on the quantum disk to the complex numbers. Assuming we have a measure $d\nu$ that invariant under the action of the generators $E,F,K$, then for any function we must impose
\begin{align}
    \int_{\mathbb{D}_q}d\nu E^m F^n h(z,z^*) = \int_{\mathbb{D}_q}d\nu \left(K^l-1\right) h(z,z^*) = 0,
\end{align}
for any $m,n,l>0$. Now suppose we want to compute the integral of an arbitrary function $h(z,z^*)$. Any such function can be expanded as
\begin{align}
    h(z,z^*) =  \sum_{k>0} h_k(y)z^k + h_0(y) + \sum_{k>0} h_{-k}(y) z^{*k}.
\end{align}
Note that the terms with extra factors of $z,z^*$,  i.e. the first and third terms in the expansion are not invariant under the action of the $K$ and can be expressed as $E^a \tilde{f}(y)$ and $F^b \hat{f}(y)$ for some functions $\tilde{f}(y), \hat{f}(y)$, and hence their integral must vanish.\footnote{The lowbrow version of this argument is to say that only the zero, or radial, mode survives the integral on the disk.} What remains is
\begin{align}
  \int_{\mathbb{D}_q}d\nu h(z,z^*) =   \int_{\mathbb{D}_q}d\nu h_0(y),
\end{align}
i.e. the integral of a purely radial function. There are two ways of making this formal expression more explicit. The first is to note that the spectrum of $y$ is given by the discrete values $q^{2n}$ for all $n\ge0$, and hence this integral can be expressed as a sum over $h_0$ evaluated on these points, namely
\begin{align}
     \int_{\mathbb{D}_q}d\nu h_0(y) = \sum_{n=0}^\infty a_n h_0(q^{2n}), \label{eq:qmeas}
\end{align}
for some coefficients $a_n$ which must be determined. This can be done by considering the (vanishing) integral of $F$ acting on $z h_0(y)$. Since 
\begin{align}
    F(zh_0(y)) = q^{-3/2} h_0(y) - q^{1/2} (1-y){h_0(q^{-2}y) - h_0( y) \over 1 - q^2 }.
\end{align}
The integral of the first and second terms on the right hand side must vanish, and so we end up with the recursion relation,
\begin{align}
 a_n = {1 \over q^{-2} - 1} \left( (1-q^{2(n+1)})a_{n+1} -  (1-q^{2n}) a_n \right) ,\quad 
  a_n = q^{-2n} a_0.
\end{align}
The value of $a_0$ is undetermined. However, if we set it to $a_0 = \pi (1-q^2)$ then we reproduce the standard integral on the classical hyperbolic disk in the limit $q \rightarrow 1$, which we will show below.

There's a more intuitive but less precise way of arriving at this result. First we need find a measure of this integral that's invariant under the action of $K,E,F$. One such measure is the following two-form
\begin{align}
    \epsilon = -{2 i} y^{-2} dz \wedge dz^*.
\end{align}
This is a two form on the disk  expressed in terms of the cotangent space element $dz \wedge dz^*$. Note that \eqref{dzr} imply we have $dz \wedge dz^* = - q^{-2} dz^* \wedge dz$.\footnote{The minus sign appears from commuting $d$ past $d$.} This form motivates the replacement
\begin{align}
    \int_{\mathbb{D}_q}d\nu = \alpha \int y^{-2} d_{q^{2}}y,
\end{align}
where $\alpha$ is some constant and the measure $d_{q^{2}}y$ stands for Jackson integral, namely that we are integrating along $y$ but in discrete steps given by the difference of $y$ at two consecutive points, i.e. $d_{q^{-2}}y = (q^{2n} - q^{2(n+1)})$. Then we have
\begin{align}
     \int_{\mathbb{D}_q}d\nu h_0(y) =  k  \sum_{n=0}^\infty h_0(q^{2n})q^{-4n}(q^{2n} - q^{2(n+1)}) =  k (1-q^2) \sum_{n=0}^\infty h_0(q^{2n})q^{-2n}.
\end{align}
This agrees with the \eqref{eq:qmeas} up to an overall constant.

An additional interesting representation of this integral uses the representation \eqref{eq:qharmrep} is
\begin{align}
     \int_{\mathbb{D}_q}d\nu h(z,z^*) = (1-q^2) \sum_{n} \lan n | h(z,z^*) | n \ran q^{-2n} = \tr\left[h(z,z^*)\frac{1}{1-z^* z}\right].
\end{align}
This will give an efficient way of computing the integrals on quantum disk.

As a check, we can compare this to the expression the classical hyperbolic disk by taking  $q \rightarrow 1$. Then we have $r^2 = |z|^2 = 1- y = 1 - q^{2n}$, and hence $q^{2n} - q^{2(n+1)} \approx 2 r dr$ and we get
\begin{align}
    \pi (1-q^2) \sum_{n=0}^\infty h_0(q^{2n})q^{-2n} = \pi\int h_0(1 - r^2) {2 r dr \over (1 - r^2)^2},
\end{align}
which is indeed the radial integral on the hyperbolic disk.

Finally, we discuss two useful relations that could be used in the computation of the integrals. The first is integration by parts that could be expressed as
\begin{gather}
    \int_{\mathbb{D}_q} d\nu E h(z,z^*) f(z,z^*) = -   \int_{\mathbb{D}_q} d\nu Kh(z,z^*) Ef(z,z^*)\,.
\end{gather}
This uses the action of $E$ on the product of two functions via coproduct and the invariance of the integral with respect to the action $E$. The second is the notion of a delta-function that can be defined as the element $\delta_{z,\xi} \in \mathbb{C}_q[z,z^*] \otimes \mathbb{C}_q[\xi,\xi^*]$ that satisfies
\begin{gather}
   f(z,z^*) = \int d\nu_\xi \delta_{z,\xi} f(\xi,\xi^*).
\end{gather}

\section{Holography}

This section will focus on holographic aspects of the quantum disk, focusing on deriving boundary anchored propagators. We will restrict our analysis to scalar fields. These fields will be arbitrary functions of $z,z^*$, and hence are elements of $\mathbb{C}_q[z,z^*]$. The formulas (\ref{efkaction1},~\ref{efkaction2},~\ref{efkaction3},~\ref{efkaction4}) will give us the action of the generators $E,F,K$ on the fields, and the Casimir (\ref{casimir},~\ref{casimirdiff}) will provide the wave equation. 

\subsection{Asymptotic analysis}

Suppose we have a field $\phi(z,z^*)$ on the  quantum disk. This field is an element $\phi \in \mathbb{C}_q[z,z^*]$. Without loss of generality, we can use the commutation relations of $z, z^*$ to express it as
\begin{align}
    \phi(z,z^*) = \sum_{k>0} \phi_k(y)z^k + \phi_0(y) + \sum_{k>0} \phi_{-k}(y) z^{*k},
\end{align}
where $y = 1 - z z^*$. We demand that the dynamics respects the $\su$ symmetry of the quantum disk, which implies that its action and equations of motion must be invariant under infinitesimal transformations generated by $K,E,F$. The simplest $\uq$ invariant wave equation linear in the field is the sum of the Casimir and mass term
\begin{align}
    \left(-\Box_q + m^2\right) \phi = 0, \quad \mathrm{with} \ \ \Box_q = C = -q^{-1}(1-z z^*)^2{\partial \over \partial z}{\partial \over \partial z^*}. \label{waveeq}
\end{align}
where we used the expression of the Casimir in \eqref{casimirdiff} in the $z,z^*$ coordinates system. This wave equation follows from the action
\begin{align}
    S =  \int_{\mathbb{D}_q}\left( -\phi \Box_q \phi + m^2 \phi^2  \right) d\nu.
\end{align}
It is instructive to see how to integrate this by parts to express the kinetic term as a square. With the appropriate measure and definition of $\Box$, we write this integral as
\begin{align}
    -\int_{\mathbb{D}_q}  \phi \Box_q \phi \, d \nu &=  q^2\int_{\mathbb{D}_q} \phi(z,z^*) y^2 {\partial \over \partial z}{\partial \over \partial z^*} \phi(z,z^*) \, y^{-2} dz \wedge dz^*, \notag\\
    &= \int_{\mathbb{D}_q} \phi(z,z^*) y^2 {\partial \over \partial z^*}{\partial \over \partial z} \phi(z,z^*) \, y^{-2} dz \wedge dz^*, \notag\\
    &= \int_{\mathbb{D}_q} \phi(z,z^*)  dz \wedge dz^* {\partial \over \partial z^*}{\partial \over \partial z} \phi(z,z^*), \notag\\
    &= \int_{\mathbb{D}_q} \left( dz^* {\partial \over \partial z^*} \phi(z,z^*)  \right)  \wedge \left( dz {\partial \over \partial z} \phi(z,z^*) \right),
\end{align}
where in the first step we commute the derivatives and pick up a factor of $q^{-2}$. In the second step, we move the measure through to the middle at no expense. In the third, we commute the two factors of the measure and  integrate by part using  $\int \bar{\partial}\left( f_1 dz f_2\right) = 0$. 

Studying the asymptotic solutions reveals some interesting aspects of fields on the quantum disk. Without any sources on the boundary, we should consider solutions that fall off near the boundary. Without loss of generality, we can consider
\begin{align}
    \phi(z,z^*) \approx \left( f(z) + f^*(z^*) \right)  \, y^{\Delta} + ... \label{falloff},
\end{align}
where we have used that $\phi^*=\phi$.
%Note that is it sufficient for the coefficient function to be only a function of $z$ since the boundary satisfies $zz^* = 1$. 
Note that on the boundary we can make the coefficient function to be only a function of $z$ since the boundary satisfies $zz^* = 1$. 
Since $y = 0$ at the boundary,  we can use equation \eqref{casimircop} to find the value of $\Delta$ as a function of the mass $m$ of the field
\begin{align}
   {(1 - q^{2 \Delta})(1 - q^{2-2\Delta}) \over (1 - q^2)^2} + m^2 = 0.
\end{align}
There are several interesting aspects of this expression. First, the left term is the eigenvalue of the Laplacian (since the solution is an eigenvector asymptotically) and was also obtained in \cite{shklyarov1998function3}. As discussed in \cite{vaksman2010quantum}, to obtain a solution with the boundary condition \eqref{falloff} that is real and non-singular everywhere in the bulk,  we must pick the principal series representation which sets $\Delta = {1 \over 2} + i \rho$. The eigenvalue becomes
\begin{align}
  \lambda_\rho= {1 + q^2 - 2 q \cos \left(2 \rho \log q\right) \over (1 - q^{2})^2} {\bf =  m^2},
\end{align}
and hence the Laplacian is bounded \cite{shklyarov1998function3,vaksman2010quantum}
\begin{align}
    {1 \over (1 + q)^2}\le \Box_q \le  {1 \over (1 - q)^2}.
\end{align}
The boundedness of the Laplacian implies the theory is finite, at least classically, and in particular has no UV divergences at coincident points. In the $q\to 1$ limit we restore the Breitenlohner-Freedman bound.

The second interesting fact about \eqref{waveeq} is that the bound on  eigenvalues places a bound on masses of (classical) fields on this space! \  For $q \approx 1$, we see that the mass is upper bounded by $\ln q$, the unit of discreteness of the radial direction, which happens to be the Planck scale from the perspective of DSSYK.

\subsection{Propagators}

 We will think of the propagator as a function on a tensor product of two copies of the quantum disk with coordinates $z,z^*$ and $w,w^*$. Hence, the propagator will be an element of the form
\begin{align}
    G(z,z^{*};w, w^*) = \sum_{i,j,k,l} A_{ij,kl} z^i z^{*j}\otimes w^k w^{*l}.
\end{align}
The task is to find solutions that satisfy
\begin{align}
    &-q^{-1} (1-zz^*)^2{\partial \over \partial z}{\partial \over \partial z^*} \phi(z,z^*) + m^2 \phi(z,z^*) = J(z,z^*), \\& \ \ \ \ \phi(z,z^*) =  \int_{\mathbb{D}_q} d\nu_w G(z,z^{*};w, w^*) J(w,w^*).
\end{align}
Similar to the classical AdS case, the bulk-to-bulk propagator on the quantum disk has to be a function of a $\us$ invariant  distance. This is a function defined on the tensor product of two quantum disks that is annihilated by the actions of $E^n,F^m, (K^k-1)$ for any $n,m,k$ acting on both coordinates. The distance is given by (the derivation could be found in \eqref{invariantd})
\begin{align}
    d(z,z^*;w, w^*) =  (1 - z z^*)^{-1} {\times_\mathrm{op}} (1 - q^{-2} z w^*) {\times_\mathrm{op}}(1 - z^* w) {\times_\mathrm{op}}(1 - w w^*)^{-1} ,
\end{align}
where the $\times_\mathrm{op}$ is the `opposite multiplication' symbol indicating that multiplication  on the first factor (the quantum disk parameterized by $z,z^*$) includes a swap: {\bf $$f(z,z^*) h(w,w^*){\times_\mathrm{op}} g(z,z^*) p(w,w^*) = g(z,z^*) f(z,z^*)  h(w,w^*) p(w,w^*), $$} where the expression on the right side of the equal sign involves standard multiplication. This means  the expanded form of the distance is
\begin{align}
    d(z,z^*;w, w^*) =  (1 - q^{-2} z w^* - z^* w + q^{-2} z^* z w^* w ) (1 - z z^*)^{-1}  (1 - w w^*)^{-1}. \label{d1}
\end{align}

This opposite multiplication is important because it takes $\us$ invariant functions and returning another invariant function, which follows from the following theorem\cite{vaksman2010quantum}.
\newtheorem{thm}{Theorem}
\begin{thm}
    If  $I_1 = \sum_i a_i \otimes b_i, \ I_2 = \sum_j c_j \otimes d_j$ are  $\us$  invariant, then $I_{12} \equiv I_1 \times_{\mathrm{op}}I_2 \equiv \sum_{i,j} c_j a_i \otimes b_i d_j$ is also $\us$ invariant.
\end{thm}

This theorem is proven in the appendix \eqref{sec:invel}. This theorem is important for us since we want to take powers of the invariant distance. The invariant distance raised to an arbitrary positive power (in the opposite multiplication sense) is given by
\begin{align}
    d^{k}  &= (1 - z z^*)^{-k} {\times_\mathrm{op}}( q^{-2k } z w^*; q^2)_k {\times_\mathrm{op}}(q^{-2(k-1)}  z^* w; q^2)_k {\times_\mathrm{op}}(1 - ww^*)^{-k},
\end{align}
where we used the $q$-Pochhammer symbol defined as
\begin{align}
    (a;q)_k = \prod_{j=0}^{k-1}(1-a q^{j}).
\end{align}
See appendix \ref{invariantd} for a derivation. It will be convenient to write the invariant distance by implementing of the opposite multiplcations as follows
\begin{align}
    d^{k}  &= (1 - z z^*)^{-k} (1 - ww^*)^{-k} \left[ ( q^{-2k } z w^*; q^2)_k {\times_\mathrm{op}}(q^{-2(k-1)}  z^* w; q^2)_k \right],
\end{align}
where the opposite multiplication takes place only within the square brackets. Inverse powers of the distance will be defined as
\begin{align}
     (d^{k})^{-1}  &=  \left[ ( q^{-2k } z w^*; q^2)_k {\times_\mathrm{op}}(q^{-2(k-1)}  z^* w; q^2)_k \right]^{-1} (1 - z z^*)^{k} (1 - ww^*)^{k}. \label{invd}
\end{align}
Note that this expression disagrees with one derived in \cite{vaksman2010quantum,shklyarov1998function3}; we found their expression to not be invariant while ours is. Thus, the general propagator is of the form
\begin{gather}
    G(z,w) =  \sum^{\infty}_{k=0} g_k \left( d^k(z,w) \right)^{-1},
\end{gather}
for some coefficients $g_k$. Only negative powers are included since the fields are required to vanish near the boundary (in the absence of sources).

\subsection{Boundary two point function}

Near the boundary we expect to have $\phi \sim  \mO(z,z^*)  (1 - z z^*)^{\Delta}$ as $(1 - zz^* )$ is taken to 0. Hence, the boundary limit of the bulk correlation function is of the form
\begin{align}
    \lan \phi(z,z^*) \phi(w,w^*) \ran = \lan \mO(z,z^*) \mO(w,w^*) \ran (1 - zz^*)^\Delta (1 - w w^*)^\Delta.
\end{align}
Thus we can read off the boundary correlation function using \eqref{invd}.  Note that the bulk coordinates are noncommutative while their boundary limits are standard commutative variables. %Hence, the boundary limit of bulk functions is a many to one map and cannot be reversed unambiguously. 
Furthermore, the commutativity of the boundary coordinates makes the $\times_\mathrm{op}$ redundant. Hence, the boundary two point function is
\begin{align}
    \lan \mO(\varphi) \mO( \theta) \ran &= ( q^{-2 \Delta }e^{i( \varphi - \theta)}; q^2)^{-1}_{\Delta}(q^{-2(\Delta-1)}   e^{-i ( \varphi - \theta)}; q^2)^{-1}_{\Delta}, \notag \\
    &=( e^{i( \varphi - \theta)}; q^2)_{-\Delta}(q^{2}   e^{-i ( \varphi - \theta)}; q^2)_{-\Delta}. \label{eq:btobdyprop}
\end{align}
As a check, the $q \rightarrow 1$ limit indeed limits to the 1-dimensional conformal two point function,
\begin{align}
    \lan \mO(\varphi)\mO( \theta)  \ran \rightarrow \left( {1 \over \sin \left[ {\theta - \varphi \over 2}\right]} \right)^{2 \Delta}.
\end{align}
As a final comment we can write down the bulk to boundary propagator. By invariance and the boundary condition $\phi \sim  \mO(z,z^*)  (1 - z z^*)^{\Delta}$, the bulk to boundary propagator must be
\begin{align}
    K_\Delta \equiv  \lan \phi(z,z^*)  \mO(w,w^*) \ran =  \left[ ( q^{-2\Delta } z w^*; q^2)_\Delta {\times_\mathrm{op}}(q^{-2(\Delta-1)}  z^* w; q^2)_\Delta \right]^{-1} (1 - z z^*)^{\Delta}, 
\end{align}
where $w = e^{i\theta}$. Amusingly, both the bulk to boundary and boundary to boundary propagators were studied by Vaksman \cite{vaksman2010quantum,shklyarov1998function3} as structures with interesting transformation properties and not motivated by the context at hand.
\section{$q$-Conformal Quantum Mechanics}

In this section we turn away from the bulk  and analyze the properties of the putative boundary quantum mechanical system with $\su$ symmetry. Note that while the action of the $\su$ on $e^{i\theta}$ at the boundary is the same as that of $SU(1,1)$ (up to constant factors) they acting differently on powers $\left( e^{i\theta} \right)^n$ where $\su$ acts with a nontrivial \cop.

Boundary primary operators are defined through the extrapolate dictionary from the expansion of bulk operators near the boundary, $\phi \sim  \mO_\Delta(e^{i \theta}) y^{\Delta}$. This expansion determines the transformation of boundary operator under $\su$ in the following way. The procedure is to act on the bulk field with a symmetry generator, e.g. $E(\phi)$ and then perform the same expansion
\begin{align}
    E(\phi) =  E[\mO(z) y^{\Delta}]   \equiv E_{\partial}[\mO(z)] y^{\Delta},
\end{align}
and hence
\begin{align}
    E_{\partial}[\mO(z)] = E[\mO(z) y^{\Delta}]  y^{-\Delta}, 
\end{align}
where $z \rightarrow e^{i \theta}$. The transformations under  $\us$ are
\begin{align}
     K_{\partial}[\mO(z)] &= \mO(q^2 z ), \label{ko} \\
     E_{\partial}[\mO(z)] &= - q^{5/2} z^2 {\mO( z ) - \mO(q^2 z ) \over z  - q^2 z }  + q^{1/2} {q^{2 \Delta} - 1 \over q^{-2}-1} z  \mO(q^2 z),\label{eo}\\
     F_{\partial}[\mO(z)] &=   q^{-3/2}  {\mO( q^{-2}z ) - \mO( z ) \over q^{-2}z  - z }  + q^{1/2} {1- q^{-2 \Delta} \over 1- q^{2}} z^{-1}  \mO( z).\label{fo}
\end{align}
These will provide Ward identities from which we can extract the correlation functions. Note that the action of the generators on the boundary satisfies the same \cop\ as that in the bulk.

The analysis of the correlation functions simplifies considerably when working in the Fourier basis where we expand local operators as
\begin{align}
    \mO(z) = \sum_{n=-\infty}^{\infty} \! \! \mO_{n} z^{-n}, \ \ \mO_n = \oint {dz\over 2 \pi i} \,  \mO(z) z^{n-1}.
\end{align}
The transformation rules \eqref{ko}, \eqref{eo},\eqref{fo} imply the Fourier modes transform as
\begin{align}
    K_{\partial}[\mO_n]&= q^{-2n} \mO_n, \label{kof} \\
     E_{\partial}[\mO_n] &=  q^{1/2}  {q^{2 \Delta - 2n - 2} - 1 \over q^{-2}-1}  \mO_{n+1} = q^{\frac12} \left[1+n-\Delta\right]_{q^{-2}} \mO_{n+1},\label{eof}\\
      F_{\partial}[\mO_n] &=    q^{1/2} {q^{2n-2}- q^{-2 \Delta} \over 1- q^{2}}  \mO_{n-1} = -q^{\frac12-2\Delta}\left[\Delta+n-1\right]_{q^2} \mO_{n-1} .\label{fof}
\end{align}
\subsection{Correlation functions}

Summetry under $\su$ implies that correlation functions are invariant under the action of any of its generators, namely that
\begin{align}
    \lan E^aF^b(K^c - 1)\left[ \mO(x_1)...\mO(x_m)\right] \ran = 0. \label{eq:mnptinv}
\end{align}
The strategy we follow in computing the position space correlation functions is by first expanding in the Fourier basis as
\begin{align}
    \lan \mO(x_1)...\mO(x_m)  \ran = \sum_{n_1,..,n_m} \lan \mO_{n_1}...\mO_{n_m} \ran x_1^{n_1} ... x_m^{n_m},
\end{align}
and then determine the coefficients $\lan \mO_{n_1}...\mO_{n_m} \ran$ by symmetry before performing the sum. Such an approach was considered before by LeClair and Bernard \cite{Bernard:1989jq} for $SL_q(2,\mathbb{R})$ case. We'll consider a few cases below for $\su$ group, the group of isometries of quantum disk.

\subsubsection{Two point function}

Our task is to determine the Fourier mode overlap $\lan \mO_m \mO_n \ran$. The simplest constraint comes the action of $K_{\partial}-1$ requiring that $\lan \mO_m \mO_n \ran = d_m \delta_{m,-n}$. We will set $d_0 = 1$. Under $E$ we get the constraint $\lan E_\partial[\mO_m] \mO_n \ran + \lan K_\partial[\mO_m] E_\partial[\mO_n] \ran = 0$, where we used the \cop\ action of $E_\partial$. Taking $m = -(n + 1)$, we obtain the recursion relation
\begin{align}
    \lan \mO_{-(n+1)} \mO_{n+1} \ran =  \alpha_n \lan \mO_{-n} \mO_{n} \ran.
\end{align}
where
\begin{align}
    \alpha_n \equiv { q^{-\Delta - n - 1} - q^{\Delta + n - 1}   \over q^{\Delta - n - 1}  - q^{-\Delta + n + 1} }.
\end{align}
This can be solved using the boundary condition $\lan \mO_0 \mO_0 \ran = 1$, and we find the Fourier coefficients
\begin{align}
    \lan \mO_m \mO_n \ran = \delta_{m,-n} q^{-2 \Delta n}\frac{(q^{2\Delta};q^2)_n}{(q^{-2\Delta+2};q^2)_n}  \label{coeff} 
\end{align}
We notice that $\braket{O_nO_{-n}} \sim q^{-2\Delta n}, n\to \infty$ and $\braket{O_n O_{-n}} \sim q^{-(2-2\Delta) n},n\to -\infty$. 
The sum can be recasted in the following compact way
\begin{gather}
     \lan \mO(x_1) \mO(x_2) \ran  = \sum_n \frac{(q^{2\Delta};q^2)_n}{(q^{-2\Delta+2};q^2)_n} \left( q^{-2\Delta} \frac{x_1}{x_2}\right)^n 
\end{gather}
And using Ramanujan $_1 \psi_1$ identity \cite{ramanujan2011notebooks} we get
\begin{gather}
     \lan \mO(x_1) \mO(x_2) \ran  = 
     \frac{(q^2;q^2)_\infty (q^{-4\Delta+2};q^2)_\infty}{(q^{-2\Delta+2};q^2)_\infty^2}\frac{(x_1/x_2;q^2)_\infty}{(q^{-2\Delta}x_1/x_2;q^2)_\infty } \frac{(q^2 x_2/x_1;q^2)_\infty}{(q^{-2\Delta+2} x_2/x_1;q^2)_\infty}
\end{gather}
Or more compactly
\begin{gather}
     \lan \mO(x_1) \mO(x_2) \ran  = 
     \frac{(q^2;q^2)_\infty (q^{-4\Delta+2};q^2)_\infty}{(q^{-2\Delta+2};q^2)_\infty^2}  ( x_1/x_2;q^2)_{-\Delta}  (q^2 x_2/x_1;q^2)_{-\Delta}
\end{gather}
Up to the overall constant factor, this is identical to the result obtained from the quantum disk in \eqref{eq:btobdyprop} after setting $x_1 = e^{i \varphi}, x_2 = e^{i \theta}$.

\subsubsection{Operator Product Expansion}
In this subsection we will discuss briefly the operator product expansion. Assume  we have two operators $\mathcal{O}_{1,2}(x)$ and that we can expand through other local operators $\mathcal{O}_i(x)$ using the rule
\begin{gather}
    \mathcal{O}_1(x) \mathcal{O}_2(y) = \sum_i \lambda^i_{12} P_{12i}(x,y,E_\partial)\mathcal{O}_i(y),
\end{gather}
The functions $P_{12i}$ roughly speaking  should give Clebsch-Gordan coefficients and should be fixed by the symmetries \cite{liskova1992clebsch,derkachev20143j,kirillov1990representations}. The coefficients $\lambda^i_{12}$ are the OPE coefficients and determined by the theory. Here we will try to compute for representation \eqref{ko},\eqref{eo}, \eqref{fo}  in the mode expansion. Thus, we want to find the coefficients $p_n$ that would give the following decomposition
\begin{gather}
    \sum_n p_n \mathcal{O}_{1,n}\mathcal{O}_{2,-n} = \mathcal{O}_{3,0}.
\end{gather}
Using asymptotics of the two-point function below \eqref{coeff}, we should demand that the coefficients should decay faster than $p_n \sim q^{(\Delta_1+\Delta_2) n}, n\to \infty$ and $p_n \sim q^{(2-\Delta_1-\Delta_2)|n|}, n\to -\infty$.

The coefficients $p_n$ could be found with the use of the following trick. Note that the operator on the right hand side is an eigenvector of the Casimir, $C_q$. Thus, acting with the Casimir on both sides of this equation gives the following recursion relation for $p_n$
\begin{gather}
q^{2+2\Delta_1}\left(q^{2\Delta_1}-q^{2n}\right)\left(q^{2\Delta_2}-q^{2n}\right)p_{n-1}+ q^{2\Delta_2}(q^{2(n+\Delta_1)}-1)(q^{2(n+\Delta_2)}-1) p_{n+1} +\notag\\ \left(q^{2(1+n+\Delta_1)} + q^{2(1+n+\Delta_2)}-
q^{2(1+\Delta_1+\Delta_2)}-q^{2(\Delta_1+\Delta_2)}- q^{2(2n+\Delta_1+\Delta_2)}-  q^{2(1+2n+\Delta_1+\Delta_2)}+ \right. \notag\\ \left.  q^{2(n+2\Delta_1+\Delta_2)}+q^{2(n+2\Delta_2+\Delta_1)}
\right) p_n = (1-q^{2\Delta_3})(1-q^{2-2\Delta_3})p_n.
\end{gather}
This is a somewhat complicated recursion relation to solve, but we considered the following natural guess for the three point function which we checked indeed satisfies this relation as well as \eqref{eq:mnptinv}. The three point function is given by
\begin{align}
    &\hspace{5cm}\braket{\mathcal{O}_1(x_1)\mathcal{O}_2(x_2)\mathcal{O}_3 (x_3)} = \notag\\
    =&(q^{2\Delta_{12,3}} x_1/x_2,q^2)_{-\Delta_{12,3}}   (q^{2\Delta_{12,3} }x_2/x_3,q^2)_{-\Delta_{23,1}}  (q^{2+2\Delta_{23,1}}x_3/x_1,q^2)_{-\Delta_{13,2}} \times\notag\\
    &(q^{2-2\Delta_{12,3}} x_2/x_1,q^2)_{-\Delta_{12,3}} (q^{2-2\Delta_{12,3}}x_3/x_2,q^2)_{-\Delta_{23,1}}   (q^{-2\Delta_{23,1}}x_1/x_3,q^2)_{-\Delta_{13,2}},
\end{align}
where $\Delta_{ij,k} = \frac{1}{2}\left(\Delta_1+\Delta_2-\Delta_3\right)$ and $\Delta_{i,j}=\Delta_i-\Delta_j$. Expanding this three point function would allow us to extract $P_{i,j,k}$.

Going back to the bulk picture, one could wonder if the three point function can be obtained from a bulk computation (it must by symmetry, but good to check explicitly). For that we just need to compute the following integral\footnote{The product of invariant elements should be taken according to the rule \eqref{sec:invel} but since $x_{1,2,3}$ are commutative elements we can just use standard multiplication.}
\begin{gather}
    G(x_1,x_2,x_3) = \int_{\mathbb{D}_q}d\nu_z K_{\Delta_1}(z,x_1)  K_{\Delta_2}(z,x_2) K_{\Delta_3}(z,x_3).
\end{gather}
$G(x_1,x_2,x_3)$ is $\us$ invariant since $K_{\Delta_{1,2,3}}$ and the integral itself are $\us$ invariant. To see that more explicitly, we notice that the integrand is an element of $\mathbb{C}_q[z,z^*] \otimes \mathbb{C}[x_1,x_2,x_3]$ and since it is $\us$ invariant we have
\begin{gather}
   0= \int_{\mathbb{D}_q}d\nu_z \Delta(E) K_{\Delta_1}(z,x_1)  K_{\Delta_2}(z,x_2) K_{\Delta_3}(z,x_3)  \notag\\
   =\int_{\mathbb{D}_q}d\nu_z \left(E_z +K_z E_{x} \right)K_{\Delta_1}(z,x_1)  K_{\Delta_2}(z,x_2) K_{\Delta_3}(z,x_3) = E_{x} G(x_1,x_2,x_3) = 0.
\end{gather}
Checking that the Witten diagram works for general dimension is complicated, but we managed to check it numerically  for negative dimensions, $\Delta_{1,2,3} = -1,-2$.

\section{Discussion}

We conclude with some outlook, remarks, and some open problems for future work.

\subsection{$q$-Conformal quantum mechanics}

It would be interesting to revisit the problem of conformal quantum mechanics but where the symmetry group is $SL_q(2)$ instead of $SL(2,\mathbb{R})$. In fact, it is simple to generalize the conformal quantum mechanical system \cite{jackiw1972introducing} to a $q$-conformal system by replacing the time derivatives with a finite difference derivative,
\begin{align}
    {\cal L} = \left[ \partial_t Q(t)\right]^2 + {g \over Q^2(t) } \rightarrow \left[ D_q Q(t)\right]^2 + {g \over Q^2(t) },
\end{align}
where 
\begin{align}
    D_{q} Q(t) = {Q(t) - Q(q^2 t) \over t - q^2 t}.
\end{align}
The operator $Q$ has dimension $-1/2$, independent of $q$. It is unclear if this is a feature or a bug, but the theory is non-local in time and perhaps makes more sense in  Euclidean time. Another unclear aspect is whether the time integral should be a regular continuous one or a  ``Jackson integral'' evaluated a discrete time steps of at powers of $q^2$. Both choices are consistent with $SL_q(2)$ symmetry.

\subsection{$q$-Schwarzian action}

One could wonder if the quantum disk is a solution to a $q$-deformed analogue of JT gravity that localizes on $q$-AdS$_2$ but with a boundary reparameterization mode. The Lagrangian of this mode would be what one would define as a $q$-deformed Schwarzian, i.e. a differential (finite difference) operator that's invariant under infinitesimal $SL_2$ transformation but with a non-trivial coproduct. See \cite{Blommaert:2023opb,Blommaert:2023wad} for proposals on the $q$-deformed Schwarzian.

The task is to find a (discrete) differential operators that's invariant under the action of $\us$, assuming we define their action on $t(u)$ to satisfy
\begin{align}
    K\Big( f[t(u)]\Big)  &= f[q^2 t(u)], \\
    E\Big( f[t(u)] \Big) &= - q^{1/2} t^2(u) {f[t(u)] - f[q^2 t(u)] \over t(u) - q^2 t(u)}, \\
    F\Big( f[t(u)] \Big) &=  q^{1/2}  {f[q^{-2} t(u)] - f[t(u)] \over q^{-2}t(u) -  t(u)}.
\end{align}
A related problem is to understand how to perform an arbitrary reparameterization of non-commutative coordinates. These 
 transformations would be generated by a $q$-deformed Virasoro algebra. Studying the central extension of this algebra would be one way of understanding the $q$-Schwarzian. We leave this for future work.

\subsection{UV lattice and Renormalization, UV/IR mixing, q affects both}

The interplay of between the discreteness of space and the renormalization group is an old and well studied problem, see for instance \cite{Collins_2004}. The upshot is that renormalization group doesn't permit Lorentz invariance to emerge in the IR if it is broken in the UV due to the existence of relevant operators that can be generated under the RG flow. A natural question is whether Lorentz invariance is preserved in the IR if the UV is invariant under a $q-$deformed version. The analogue for the quantum disk is $SU_q(1,1)$ in the UV and $SU(1,1)$ in the IR.

One potential hurdle is the problem of UV/IR mixing associated to non-commutative spaces, discussed for instance here \cite{Seiberg:1999vs}. In fact, the quantum disk has a hint of this in the spectrum of the Laplacian,
\begin{align}
    (1 + q)^{-2} \le \Box_q\le (1-q)^{-2},
\end{align}
where both the IR and UV bounds are controlled by the same paramter $q$. One would need to compute loop diagrams to truly settle this.

\subsection{q-Conformal Blocks}
Using the ideas similar to the usual conformal field theory we can use the operator product expansion and assosiativity to constrain the possible $\su$ invariant field theories. For that, we must to study the four-point functions
\begin{gather}
    \braket{O_1(z_1) O_2(z_2) O_3(z_3) O_4(z_4)}.
\end{gather}
As in the case of the usual CFTs we can not constrain the possible form of this expression, but we can expand it in terms of conformal blocks that satisfy the following equation
\begin{gather}
    C_q G_\Delta(z_1,z_2,z_3,z_4) = C_\Delta G_\Delta(z_1,z_2,z_3,z_4).
\end{gather}
It would be interesting to study the possible solutions of this equation.

\subsection{DSSYK at $q<1$ and large $\beta$}

One of the motivations of this work was to study a toy model that resembles the putative bulk dual of DSSYK since it too has a noncommutative nature. A natural question is whether the quantum disk emerges from some regime of DSSYK.

There's some evidence that this might be the case. It was noted in \cite{Lin:2023trc} that invariance under $\us$ emerges in the low temperature limit of DSSYK at $q<1$. This symmetry fixes the form of the two point function. The exact expression of the two point function was found in \cite{Berkooz:2018jqr}, so it is simply a matter of taking its low temperature limit. We leave this for future work.

\section*{Acknowledgements}

We thank    Micha Berkooz, Mikhail Isachenkov, Henry Lin, Simon Lin, Ohad Mamroud, Alexei Milekhin and Yifan Wang for insightful discussion and correspondence.
F.K.P.
is currently a Simons Junior Fellow at New York University and supported
by a grant 855325FP from the Simons Foundation.

\appendix
\section{Introduction to Quantum Groups}
\label{sec:appa}
\subsection{General Philosophy}
This appendix contains what hopefully will be an accessible and pedagogical review of quantum groups. For that, we will start with the general philosophy or idea that lead to the creation of  quantum groups. Usually when we study groups or other classical manifolds we have an idea that it is some sort of geometric object that could be thought as some surface in higher dimensional flat space.  Instead, it is more convenient here to use the language of algebraic geometry which analyzes functions on the manifold whose structures are reflected in the algebra of functions. 
Thus, if we have a topological space $\mathcal{M}$ we will denote $\mathcal{O}(\mathcal{M})$ as a set of all functions $f \in \mathcal{O}(\mathcal{M}), f:\mathcal{M} \to \mathbb{C}$ on the given topological manifold. This set automatically has the structure of commutative algebra. We have the following operations and special element $\mathbf{1}$
\begin{align}
& \forall f_{1,2} \in \mathcal{O}(\mathcal{M}), \quad (f_1 + f_2)(x) = f_1(x) + f_2(x), \quad (f_1 \cdot f_2)(x) = f_1(x) f_2(x) \quad \forall x \in \mathcal{M}\,, \notag\\
&\mathbf{1}(x) = 1, \quad\forall x \in \mathcal{M},\quad  \mathbf{1}\cdot f = f
\label{eq:alg}
\end{align}
it is easy to check that $\mathbf{1}$ is indeed a unit in the algebra $\mathcal{O}(\mathcal{M})$. By studying the algebraic properties of this commutative algebra we can understand the geometric properties of the topological manifold $\mathcal{M}$. For instance, the ideals of $\mathcal{O}(\mathcal{M})$ will correspond to  submanifolds and so on.
\subsection{Hopf algebra} \label{hopf}
Now assume that $\mathcal{M}$ is not only just a topological manifold but also a group. Then the algebra of functions $\mathcal{O}(\mathcal{M})$ gets additional structures that encode the group structure of this topological group. Thus, we have now a special element $e \in \mathcal{M}$, the identity, and the two additional operations, inverse and product. Evaluating a function on the manifold on the element $e$ gives  a %The existence of the special element means that we can evaluate our functions at given element. That evaluation could be thought as a
map from $\mathcal{O}(\mathcal{M})$ to complex functions. Such operations we will denote as the \cou
\begin{gather}
    \eta: \mathcal{O}(\mathcal{M}) \to \mathbb{C}, \quad \eta(f) = f(e).\label{eq:counit}
\end{gather}
The existence of a group product means we can map any function of a single variable to  a function of two variables
\begin{gather}
    \Delta: \mathcal{O}(\mathcal{M}) \to \mathcal{O}(\mathcal{M}) \otimes \mathcal{O}(\mathcal{M}), \quad \Delta(f)(g_1,g_2) = f(g_1 g_2),\quad \forall g_{1,2} \in \mathcal{M} \label{eq:coproduct}
\end{gather}
where we assumed  we can make a replacement $\mathcal{O}(\mathcal{M} \times \mathcal{M}) = \mathcal{O}(\mathcal{M}) \otimes \mathcal{O}(\mathcal{M})$. The map $\Delta$ is the {\bf coproduct}. Associativity of a group product suggests a coassociativy of the \cop\ $(\Delta \otimes {\rm id}) \Delta =({\rm id }\otimes \Delta) \Delta$. The group axioms imply that the following relation must hold
\begin{gather}
    ({\rm id} \otimes \eta)\Delta = (\eta \otimes {\rm id})\Delta = {\rm id}\,. \label{eq:costr}
\end{gather}

If a set has a \cou  \ $\eta$ and \cop\ then it is a {\bf coalgebra}. If an algebra has the structure of a {\bf coalgebra} and these structures are compatible in the sense 
\begin{gather}
    \Delta(ab) = \Delta(a) \Delta(b), \quad \eta(ab) = \eta(a)\eta(b)\,, \label{eq:bistr}
\end{gather}
then the set is a {\bf bialgebra}.

The inverse allows to define an additional structure that maps the algebra of functions back to itself
\begin{gather}
    S: \mathcal{O}(\mathcal{M}) \to \mathcal{O}(\mathcal{M}), \quad S(f)(g) = f(g^{-1}), \quad \forall g \in \mathcal{M}\,, \label{eq:antipodef} 
\end{gather}
where the operation $S$ is the \anti. The group axioms imply that the following relation must hold
\begin{gather}
    m(S \otimes {\rm id}) \Delta =    m(S \otimes {\rm id}) \Delta = \eta\,. \label{eq:antipodrel}
\end{gather}

Finally, if we have a { \bf bialgebra} with the \anti \ $S$ satisfying the above relation then the {\bf bialgebra} is a {\bf Hopf algebra}. We see that the algebra of functions on any topological group satisfies these relations. This results in a commutative {\bf Hopf algebras}.\footnote{However, it is not {\bf cocommutative} because $f(g_1 g_2) \neq f(g_2 g_1)$ for general $f$.} This defines a group. On the other hand, a quantum group is defined by a  non-commutative {\bf Hopf algebra}. 

\subsubsection{Hopf algebra $\mathcal{O}(SL_2(\mathbb{R}))$}
Before proceeding to the quantum case, let us consider first the most simple classical group $SL_2(\mathbb{R})$ and see how the above definitions work. A group $SL_2(\mathbb{R})$ could be defined as follows
\begin{gather}
    SL_2(\mathbb{R}) = \left\{\begin{pmatrix}
    t_{11} & t_{12}\\
    t_{21} & t_{22}
    \end{pmatrix}\in \mathbb{R}^4: t_{11}t_{22}-t_{12}t_{21} = 1
    \right\}\,,
\end{gather}
the elements of the matrix $t_{ij}$ could be thought of as coordinates in the sense they are maps from group elements $g$ to their matrix elements $t_{ij}(g)$ in the above two dimensional representation, and they furnish a basis for functions on the group. Hence, the algebra of functions is the algebra of all possible polynomials in commutative variables $t_{ij}$ subject to the relation $t_{11}t_{22}-t_{12}t_{21} = 1$. This  set is denoted by
\begin{gather}
    \mathcal{O}(SL_2(\mathbb{R})) = \mathbb{C}[t_{11},t_{12},t_{21},t_{22}]/\left(t_{11}t_{22}-t_{12}t_{21}-1\right)\,.
\end{gather}
The structures $\eta$, $S$ and $\Delta$ are determined by the group structure.
Roughly speaking, given the coordinates of a group element, these operations compute the coordinates of the unit, product and inverse of this group element
\begin{gather}
    \eta(t_{ij}) = \delta_{ij}, \quad \Delta(t_{ij}) = t_{ik}\otimes t_{kj}, \quad S(t_{ij}) = \epsilon_{ik}\epsilon_{jl}t_{kl}\,.
\end{gather}
Let us expand briefly on the \cop\ here. If we have two elements $g,h$ with coordinates $t_{ij}(g),t_{lm}(h)$ than the \cop\  gives the coordinates of their product
\begin{gather}
    \Delta(t_{ij})(g,h) = t_{ik}(g) t_{kj}(h) = g_{ik} h_{kj} = t_{ij}(gh)\,,
\end{gather}
and one can check that these operations satisfy the relations \eqref{eq:alg}-\eqref{eq:antipodrel} giving us a {\bf Hopf algebra} $\mathcal{O}(SL_2(\mathbb{R}))$. Usually, along with the group it is useful to consider a universal enveloping algebra $U(\mathfrak{sl}_2)$ that could be thought as an algebra of the vector fields generated by the left action of the group. Thus by considering the infinitesimal transformation we find three vector fields $E,F,H$ that are subject to the following relations
\begin{gather}
    [H,E] = 2 E, \quad [H,F] = - 2 F, \quad [E,F]=H\,.
\end{gather}
 $U(\mathfrak{sl}_2)$ is a free generated algebra in variables $E,F,H$ subject to the above relations.
 
Since we are adopting an algebraic approach it would be useful to understand how to derive this algebra starting from Hopf algebra $\mathcal{O}(SL_2(\mathbb{R}))$. For that we notice first that $U(\mathfrak{sl}_2)$ is a {\bf Hopf algebra} where the {\bf coalgebra} structure is defined as
\begin{gather}
    \eta(J_i) = 0, \quad \Delta(J_i) = 1 \otimes J_i + J_i \otimes 1,\quad S(J_i) = - J_i,\quad J_i = E,F,H\label{eq:usl2co}\,,
\end{gather}
then we notice that since $E,F,H$ are  vector fields and have a natural action on elements of the algebra $\mathcal{O}(SL_2(\mathbb{R}))$ then when combined with \cou\ we can define a pairing between these two { \bf Hopf algebras}
\begin{gather}
    \sum_{ijk}\langle c_{ijk}E^iF^jH^k,f\rangle = \sum c_{ijk}\eta(E^i F^jH^kf),\quad \sum c_{ijk} E^i F^j H^k \in U(\mathfrak{sl}_2), f \in \mathcal{O}(SL_2(\mathbb{R})
\end{gather}
where we apply consequently vector fields $E,F,H$ on the given function $f \in \mathcal{O}(SL_2(\mathbb{R}))$.  This pairing is the key to  understanding of the relation between $U(\mathfrak{sl}_2)$ and $\mathcal{O}(SL_2(\mathbb{R}))$.  First of all, we notice that this pairing allows to restore the action of $U(\mathfrak{sl}_2)$ on $\mathcal{O}(SL_2(\mathbb{R}))$. Indeed $J(f) = (\langle J,\cdot \rangle \otimes {\rm id}) \Delta(f)$. Second of all, this pairing respects the structure of the { \bf Hopf algebras} namely
\begin{gather}
    \langle JK,f\rangle = (\langle J,\cdot\rangle \otimes \langle F,\cdot \rangle) \Delta(f), \quad  \langle J,f g\rangle = (\langle \cdot,f\rangle \otimes \langle \cdot,g \rangle) \Delta(J) \notag\\
    \langle 1,f\rangle = \eta(f), \quad \langle J, 1 \rangle = \eta(J),\quad f,g \in \mathcal{O}(SL_2(\mathbb{R})), J,K \in U(\mathfrak{sl}_2)\,.
\end{gather}
Now, if we have two { \bf Hopf algebras} $U,V$ that have a pairing $\langle,\rangle$ we would say these two {\bf Hopf algebras} are dual to each other. We then define that $U_q(\mathfrak{sl}_2)$ is a dual algebra of $\mathcal{O}(SL_2(\mathbb{R}))$. This same approach we will adopt for quantum algebras.

\subsubsection{Hopf algebra $\mathcal{O}(SL_2(\mathbb{R}))$} \label{idderiv}
Having understood how from classical groups we can get the { \bf Hopf algebra}, we can forget about our base manifold and start immediately with some algebra. Thus we consider the following algebra
\begin{gather}
    \mathcal{O}_q(SL_{2}(\mathbb{R})) =  \mathbb{C}[t_{11},t_{12},t_{21},t_{22}]/I, \notag\\
    I = \left(t_{11}t_{12}-qt_{12}t_{11}, t_{11}t_{21}-q t_{21} t_{11}, t_{12}t_{22} - q t_{22} t_{12}, t_{21}t_{22} - q t_{22} t_{21}, t_{12}t_{21} = t_{21}t_{12}, \right. \notag\\
    \left.t_{11}t_{22} - t_{22}t_{11} = (q-q^{-1}) t_{12}t_{21}, t_{11}t_{22} - q t_{12} t_{21} -1\right), \label{eq:quantmat}
\end{gather}
and polynomials generated by non-commutative variables $t_{ij}$. The structure of the { \bf coalgebra} is defined as
\begin{gather}
    \eta(t_{ij}) = \delta_{ij}, \quad \Delta(t_{ij}) = t_{ik}\otimes t_{kj}, \quad S(t_{ij}) = \epsilon_{il}\epsilon_{jm} q^{-\epsilon_{ij}} t_{lm}\,, 
\end{gather}
from this we see that the quantum group as a group has a structure that resembles the usual $SL_2(\mathbb{R})$ group with usual group product and identity. So as a group a quantum group is no different from any other group but the only difference is that coordinates are not commutative. 

Now let us try to use the above procedure to construct the $U_q(\mathfrak{sl}_2)$. For that we will define the following elements $E,F,K$ through the pairing, such that their pairing with basis elements is defined as
\begin{gather}
   q^{-1} \langle K,a \rangle =  q \langle K,d \rangle = q^{1/2} \langle E, b \rangle = q^{-1/2}\langle F, c \rangle = 1\,,
\end{gather}
and the action on the product of two elements satisfies the following relation
\begin{align}
    &\langle E, ab \rangle = \langle E,a \rangle \eta(b) + \langle K,a\rangle \langle E,b\rangle,\notag\\
    &\langle F, ab \rangle = \langle F,a\rangle \langle K^{-1},b\rangle + \eta(a)\langle F,b\rangle,\notag\\
    &\langle K,ab\rangle = \langle K,a\rangle \langle K,b\rangle\,.
\end{align}
one can check that such a definition is compatible with the ideal $I$, and therefore defines a unique pairing on all elements of $\mathcal{O}_q(SL_2(\mathbb{R}))$. 

This pairing can be used to establish the following ideals of $\uq$
\begin{gather}
    I_E = K E - q^2 E K, \quad  I_F = K^{-1} F - q^{2} F K^{-1} \,, 
\end{gather}
namely that $\langle I_{E,F} , t_{ij} \rangle = 0$ as we show below. Let us concentrate on $I_E$.  Now if we have a product of two elements we can rewrite the action of $I_E$ using the \cop 
\begin{gather}
    \Delta(a) = a'_i \otimes a''_i, \quad \Delta(b) = b'_j \otimes b''_j, \notag\\
    \Delta(ab) = a'_i b'_j \otimes a''_i b''_j\,.
\end{gather}
where $a,b$ (including the indexed and primed) $\in \osq$.  These give
\begin{gather}
    \langle K E,ab\rangle = \langle K, a'_i b'_j \rangle  \langle E, a''_i b''_j \rangle =\langle K, a'_i  \rangle \langle K,  b'_j \rangle  \left( \langle E, a''_i\rangle \eta(b''_j) + \langle K, a''_i\rangle \langle E, b''_j\rangle \right)  = \notag\\
    \langle K E , a  \rangle \langle K,  b \rangle + \langle K, a  \rangle \langle KE,  b \rangle
\end{gather}
and
\begin{gather}
    \langle E K,ab\rangle = \langle E, a'_i b'_j \rangle  \langle K, a''_i b''_j \rangle =  \left( \langle E, a'_i\rangle \eta(b'_j) + \langle K, a'_i\rangle \langle E, b'_j\rangle \right) \langle K, a''_i  \rangle \langle K,  b''_j \rangle  \ = \notag\\
    \langle  E K , a  \rangle \langle K,  b \rangle + \langle K, a  \rangle \langle EK,  b \rangle
\end{gather}
 eventually yielding
\begin{gather}
    \langle I_E, ab\rangle = \langle I_E,a \rangle \langle K,b\rangle +  \langle I_E,b \rangle \langle K,a\rangle,
\end{gather}
and therefore it is easy to see that $I_E \equiv 0$ (similarly for $I_F \equiv 0$). 
Analogously, another ideal is
\begin{gather}
    P = EF - FE - \frac{K - K^{-1}}{q-q^{-1}}\,.
\end{gather}
Let us try to compute the pairing of this element with any element $a,b,c,d$. It is easy to check that
\begin{gather}
    \langle P,a\rangle =  \langle P,b\rangle = \langle P,c\rangle = \langle P,d\rangle = 0\,,
\end{gather}
and similarly all products $ab$.  The most simple action is that of $K$
\begin{gather}
     \left\langle \frac{K - K^{-1}}{q-q^{-1}}, a b\right\rangle = \frac{\langle K, a\rangle \langle K, b\rangle - \langle K^{-1}, a\rangle \langle K^{-1}, b\rangle }{q-q^{-1}} = \notag\\
     =  \left\langle \frac{K - K^{-1}}{q-q^{-1}}, a \right\rangle  \left\langle K^{-1}, b \right\rangle +  \left\langle K, a \right\rangle \left\langle \frac{K - K^{-1}}{q-q^{-1}}, b \right\rangle  
\end{gather}
Then the action of $EF$ on those two elements is 
\begin{gather*}
    \langle EF,ab\rangle = \langle E, a'_i b'_j\rangle \langle F, a''_i b''_j \rangle = 
    \left( \langle E, a'_i\rangle \eta(b'_j) + \langle K, a'_i\rangle \langle E, b'_j\rangle \right)   \left(  \langle F, a''_i \rangle   \langle K^{-1}, b''_j \rangle  + \eta(a''_i) \langle F, b''_j \rangle \right) = \notag\\
    \langle E, a'_i\rangle  \langle F, a''_i \rangle   \langle K^{-1}, b \rangle  + \langle E, a\rangle \langle F, b \rangle +\langle K, a'_i\rangle \langle E, b'_j\rangle \langle F, a''_i \rangle   \langle K^{-1}, b''_j \rangle + \langle K, a\rangle \langle E, b'_j\rangle \langle F, b''_j \rangle = \notag\\
      \langle E F, a\rangle \langle K^{-1}, b \rangle  + \langle E, a\rangle \langle F, b \rangle +\langle K F, a\rangle \langle EK^{-1}, b\rangle  + \langle K, a\rangle \langle EF, b\rangle 
    \,,
\end{gather*}
Analogously for the opposite ordering
\begin{gather*}
    \langle FE,ab\rangle = \langle F, a'_i b'_j\rangle \langle E, a''_i b''_j \rangle = 
    \left(  \langle F, a'_i \rangle   \langle K^{-1}, b'_j \rangle  + \eta(a'_i) \langle F, b'_j \rangle \right)   
    \left( \langle E, a''_i\rangle \eta(b''_j) + \langle K, a''_i\rangle \langle E, b''_j\rangle \right)    = \notag\\
    \langle F, a'_i \rangle   \langle E, a''_i\rangle  \langle K^{-1}, b\rangle  +  \langle F, a'_i \rangle   \langle K^{-1}, b'_j \rangle \langle K, a''_i\rangle \langle E, b''_j\rangle   + \langle F, b \rangle \langle E, a\rangle  + \langle F, b'_j \rangle \langle K, a\rangle \langle E, b''_j\rangle = \notag\\
    \langle FE, a \rangle   \langle K^{-1}, b\rangle  +  \langle FK, a \rangle   \langle K^{-1}E, b\rangle  + \langle F, b \rangle \langle E, a\rangle  + \langle FE , b \rangle \langle K, a\rangle\,,
\end{gather*}
where we have used the properties of the \cop\ $\Delta$ and \cou \ $\eta$.  Combing the two above relations we  get
\begin{gather}
      \langle P,ab\rangle =
      \langle P ,a \rangle \langle K^{-1},b\rangle + \langle K,a \rangle    \langle P,b \rangle ,
\end{gather}
therefore $P \equiv 0$.   Together with $I_{E,F}$, these provide the definition of quantum universal enveloping algebra $U_q(\mathfrak{sl}_2)$.
\subsection{Invariant elements} \label{sec:invel}
An important notion is that of invariant elements. To be more precise assume  we have two modules $\mathcal{M}, \mathcal{N}$ of $\us$. An element $d \in \mathcal{M} \otimes \mathcal{N}$ is invariant if
\begin{gather}
    E(d) = F(d) = K(d)-d = 0, \quad d=\sum_i m_i \otimes n_i,\notag
\end{gather}
where for example
\begin{align}
    E(d) = \sum_i E(m_i)\otimes n_i + K(m_i)  \otimes E(n_i) = 0\,.
\end{align}

If we assume that $\mathcal{M},\mathcal{N}$ are algebras and we have two invariant elements $d_{1,2}$ then, as discussed in \cite{2008arXiv0803.3769V}, we can construct a new invariant element $d_3$ in the following way
\begin{gather}
    d_{1,2} = \sum_i m^{1,2}_i \otimes n^{1,2}_i, \quad d_3 = \sum_{i,j} m^1_i m^2_j \otimes n^2_j n^1_i\,,
\end{gather}
the straightforward computation shows that this element is indeed invariant
\begin{gather}
    E(d_3) = \sum_{i,j} \left[ E(m^1_i) m^2_j \otimes n^2_j n^1_i +  K(m^1_i) E(m^2_j) \otimes n^2_j n^1_i + K(m^1_i) K(m^2_j) \otimes E(n^2_j) n^1_i  \right.\notag\\ \left. +  K(m^1_i) K(m^2_j) \otimes K(n^2_j) E(n^1_i) \right] = \notag\\
    \sum_i (K(m_i^1) \otimes 1) E(d_2) (1\otimes n_i^1) + \sum_{i,j} (1\otimes n^2_j)  E(d_1) \left(m^2_j \otimes 1\right) = 0
\end{gather}
\subsection{Powers of the invariant distance} \label{invariantd}

The steps of this proof follow that of \cite{shklyarov1998function3} but differs in detail. Consider the tensor product of two quantum disks paramterized by $t_{ij}$ and $\tau_{ij}$. The following linear combinations of $\tau_{ij} \otimes t_{kl}$ are invariant under the action of $\us$
\begin{align}
    A &= t_{22}\tau_{11} - q^{-1}t_{12}\tau_{21}, \\
    B &= t_{11}\tau_{22} - qt_{21}\tau_{12},
\end{align}
where we have dropped the tensor product symbol for brevity. The disk coordinates are related to these variable in the following way
\begin{align}
    z &= q^{-1} t_{21}^{-1} t_{11}, \ z^* = t_{22} t^{-1}_{12} \\
    w &= q^{-1} \tau_{21}^{-1} \tau_{11}, \ w^* = \tau_{22} \tau^{-1}_{12}.
\end{align}

The two invariants can be expressed in terms of the disk coordinates as follow while implementing the opposite product on the first tensor factor as
\begin{align}
    A &= (- q^{-1}t_{12}\tau_{21}) \times_\mathrm{op}(1 - q^2 z^* w), \\
    B &= (1 - z w^*)\times_\mathrm{op}(- qt_{21}\tau_{12}).
\end{align}
Using the following relations that follow from the algebra \eqref{coordsl2}
\begin{align}
    t_{12} \tau_{21}  \times_\mathrm{op}(z^* w) = q^{-2} (z^* w) \times_\mathrm{op} \tau_{21} t_{12}, \ (z w^*)  \times_\mathrm{op}t_{21} \tau_{12} = q^{-2} t_{21} \tau_{12}  \times_\mathrm{op}(z w^*),
\end{align}
the powers of the invariants are 
\begin{align}
    A^k &= (- q^{-1}t_{12}\tau_{21})^k  \times_\mathrm{op}(q^2 z^* w;q^2)_k, \\
    B^k &= (z w^*; q^2)_k  \times_\mathrm{op} (- qt_{21}\tau_{12})^k.
\end{align}
Taking their opposite product we get
\begin{align}
    B^k A^k &= (z w^*; q^2)_k  \times_\mathrm{op} (- qt_{21}\tau_{12})^k  \times_\mathrm{op}(- q^{-1}t_{12}\tau_{21})^k  \times_\mathrm{op}(q^2 z^* w;q^2)_k \\
    &= q^{-2m}(z w^*; q^2)_k  \times_\mathrm{op}(1-zz^*)^{-k} (1-w w^*)^{-k}  \times_\mathrm{op}(q^2 z^* w;q^2)_k
\end{align}
where we used $t_{12}t_{21} = - q^{-1}(1 - z z^*)^{-1}$ and similarly for $w$. Commuting the middle factors across gives
\begin{align}
    B^k A^k = q^{-2k}  (1-zz^*)^{-k}   \times_\mathrm{op} (q^{-2k} z w^*; q^2)_k  \times_\mathrm{op}  (q^{-2(k-1)} z^* w;q^2)_k  \times_\mathrm{op}(1-w w^*)^{-k}
\end{align}
The extra factors of $q$ inside the $q$-Pochhammers comes from commuting the $(1-z z^*)$ and $(1-w w^*)$ through. For $k=1$ this is just the invariant distance $d$ in \eqref{d1}. This constructs the appropriate invariant power of $d$.

\section{$R$ Matrix approach to $\sq$} \label{rmatrix}
The non-commutative structure \eqref{slq} defines a specific quantum deformation of the $SL(2)$. These relations are self consistent, and any polynomial of the matrix elements can be written as a sum over monomials with a given ordering of the elements $a,b,c,d$ without changing the degree of the polynomial. These two properties are guaranteed by the Yang-Baxter equation
\begin{align}
    g^i_{\ k} \, g^{j}_{\ l} \, R^{kl}_{\ \ m n} = R^{ji}_{\ \ rs}\,  g^{s}_{\ n} \, g^r_{\ m}, \label{ggr}
\end{align}
where $g \in \sq$ and $R$ is a $4\times4$ matrix of complex numbers satisfying the Yang-Baxter (YB) equation
\begin{align}
    R_{12}R_{23}R_{12} = R_{23} R_{12} R_{23}, \label{eq:yba}
\end{align}
where $R_{12} = R^{i_1 i_2}_{\ \ j_1 j_2} \delta^{i_3}_{\ j_3}$ and the others defined similarly. The flipped indices on the right hand side of \eqref{ggr} is intentional, giving a relation between the two different orders of all the matrix elements. \footnote{This equation is sometimes written as $g^i_{\ k} \, g^{j}_{\ l} \, \hat{R} ^{kl}_{\ \ m n} = \hat{R}^{ji}_{\ \ rs}\,  g^{r}_{\ m} \, g^s_{\ n}$ where $\hat{R} = R \circ \tau$ and $\tau$ is a swap.} For the case at hand we have
\begin{align}
     \begin{pmatrix}
aa & ab & ba & bb\\
ac & ad & bc & bd\\
ca & cb & da & db\\
cc & cd & dc & dd
\end{pmatrix} 
 \begin{pmatrix}
q & 0 & 0 & 0\\
0 & 1 & q-q^{-1} & 0\\
0 & 0 & 1 & 0\\
0 & 0 & 0 & q
\end{pmatrix}  = \begin{pmatrix}
q & 0 & 0 & 0\\
0 & 1 & q-q^{-1} & 0\\
0 & 0 & 1 & 0\\
0 & 0 & 0 & q
\end{pmatrix} \begin{pmatrix}
aa & ba & ab & bb\\
ca & da & cb & db\\
ac & bc & ad & bd\\
cc & dc & cd & dd
\end{pmatrix} 
\end{align}

The definition of commutation relations \eqref{ggr} allows the definition of the space where the elements of the quantum groups act. Thus, we introduce the coordinate functions $x_i$ that satisfy the relation
\begin{gather}
    R^{kl}_{mn} x_k x_l = x_m x_n,
\end{gather}
then one can check that if we introduce new objects $y_k=g^i_k x_i$ they also satisfy the same relation.

On general grounds, the solution of the YB equation is quite complicated, but Drinfeld managed to find a universal solution when $R$ acts on a tensor product of the representations of quantum group \cite{drinfeld1986quantum}. For the case at hand where we have $V_1\otimes V_2$ with some action of $\us$ then the $R$ matrix is given by
\begin{gather}
    R = \exp_{q^2} \left((q^{-1}-q) E\otimes F\right)q^{-\frac{H\otimes H}{2}}
\end{gather}

\section{Casimir and derivatives} \label{app:casandder}
 Here we show by explicit computation that 
 \begin{align}
     C_q = q^{-1}\left(1-z z^*\right)^2 \frac{\partial}{\partial z} \frac{\partial}{\partial z^*}
 \end{align}
 Consider the action of Casimir operator on the following monomial
 \begin{gather}
     C_q \left(z^n z^{*m}\right) = \notag\\
      q^3 [n]_{q^2} [m]_{q^{2}}z^{n+1} z^{*m+1} + q[n]_{q^{2}} [m]_{q^{2}} z^{n-1} z^{*m-1} -q(1+q^2)[n]_{q^2}[m]_{q^2} z^n z^{*m}= \notag\\
      q [n]_{q^2}[m]_{q^2} z^{n-1}\left(1-(1+q^2) z z^* + q^2 z^2 z^{*2}\right) z^{*m-1}
 \end{gather}
Then we compute the action of the differential operator on the same monomial,
\begin{gather}
    q^{-1}\left(1-z z^*\right)^2 \frac{\partial}{\partial z} \frac{\partial}{\partial z^*}(z^n z^{*m}) = q^{-3}[m]_{q^{2}} [n]_{q^{2}} z^{n-1} y^2 z^{*m-1}  = \notag\\
    q[m]_{q^2}[n]_{q^2}z^{n-1}\left(1 - (1+q^2)z z^* +q^2 z^2 z^{*2}\right)z^{*m-1}=C_q\left(z^n z^{*m}\right)
\end{gather}
These two expressions agree. This completes the proof since these monomials span the vector space $\mathbb{C}_q[z,z^*]$.

\bibliographystyle{ourbst}
\bibliography{SL2q.bib}

\providecommand{\href}[2]{#2}\begingroup\raggedright\begin{thebibliography}{10}

\bibitem{Seiberg:1999vs}
N.~Seiberg and E.~Witten, {{String theory and noncommutative geometry}},
  \href{http://dx.doi.org/10.1088/1126-6708/1999/09/032}{JHEP {\bf 09}, 032,
  1999},
  [\href{http://arxiv.org/abs/arXiv:hep-th/9908142}{{arXiv:hep-th/9908142}}].

\bibitem{Connes:1997cr}
A.~Connes, M.~R. Douglas and A.~S. Schwarz, {{Noncommutative geometry and
  matrix theory: Compactification on tori}},
  \href{http://dx.doi.org/10.1088/1126-6708/1998/02/003}{JHEP {\bf 02}, 003,
  1998},
  [\href{http://arxiv.org/abs/arXiv:hep-th/9711162}{{arXiv:hep-th/9711162}}].

\bibitem{Gross:2000ph}
D.~J. Gross and N.~A. Nekrasov, {{Dynamics of strings in noncommutative gauge
  theory}}, \href{http://dx.doi.org/10.1088/1126-6708/2000/10/021}{JHEP {\bf
  10}, 021, 2000},
  [\href{http://arxiv.org/abs/arXiv:hep-th/0007204}{{arXiv:hep-th/0007204}}].

\bibitem{Gross:2000ss}
D.~J. Gross and N.~A. Nekrasov, {{Solitons in noncommutative gauge theory}},
  \href{http://dx.doi.org/10.1088/1126-6708/2001/03/044}{JHEP {\bf 03}, 044,
  2001},
  [\href{http://arxiv.org/abs/arXiv:hep-th/0010090}{{arXiv:hep-th/0010090}}].

\bibitem{Gross:2000wc}
D.~J. Gross and N.~A. Nekrasov, {{Monopoles and strings in noncommutative gauge
  theory}}, \href{http://dx.doi.org/10.1088/1126-6708/2000/07/034}{JHEP {\bf
  07}, 034, 2000},
  [\href{http://arxiv.org/abs/arXiv:hep-th/0005204}{{arXiv:hep-th/0005204}}].

\bibitem{Alvarez-Gaume:2003lup}
L.~Alvarez-Gaume and M.~A. Vazquez-Mozo, {{General properties of noncommutative
  field theories}},
  \href{http://dx.doi.org/10.1016/S0550-3213(03)00582-0}{Nucl. Phys. B {\bf
  668}, 293--321, 2003},
  [\href{http://arxiv.org/abs/arXiv:hep-th/0305093}{{arXiv:hep-th/0305093}}].

\bibitem{Berkooz:2018jqr}
M.~Berkooz, M.~Isachenkov, V.~Narovlansky and G.~Torrents, {{Towards a full
  solution of the large N double-scaled SYK model}},
  \href{http://dx.doi.org/10.1007/JHEP03(2019)079}{JHEP {\bf 03}, 079, 2019},
  [\href{http://arxiv.org/abs/arXiv:1811.02584}{{arXiv:1811.02584 [hep-th]}}].

\bibitem{Berkooz:2018qkz}
M.~Berkooz, P.~Narayan and J.~Simon, {{Chord diagrams, exact correlators in
  spin glasses and black hole bulk reconstruction}},
  \href{http://dx.doi.org/10.1007/JHEP08(2018)192}{JHEP {\bf 08}, 192, 2018},
  [\href{http://arxiv.org/abs/arXiv:1806.04380}{{arXiv:1806.04380 [hep-th]}}].

\bibitem{Berkooz:2022mfk}
M.~Berkooz, M.~Isachenkov, M.~Isachenkov, P.~Narayan and V.~Narovlansky,
  {{Quantum groups, non-commutative AdS$_{2}$, and chords in the double-scaled
  SYK model}}, \href{http://dx.doi.org/10.1007/JHEP08(2023)076}{JHEP {\bf 08},
  076, 2023}, [\href{http://arxiv.org/abs/arXiv:2212.13668}{{arXiv:2212.13668
  [hep-th]}}].

\bibitem{Lin:2023trc}
H.~W. Lin and D.~Stanford, {{A symmetry algebra in double-scaled SYK}},
  \href{http://dx.doi.org/10.21468/SciPostPhys.15.6.234}{SciPost Phys. {\bf
  15}, 234, 2023},
  [\href{http://arxiv.org/abs/arXiv:2307.15725}{{arXiv:2307.15725 [hep-th]}}].

\bibitem{Cotler:2016fpe}
J.~S. Cotler, G.~Gur-Ari, M.~Hanada, J.~Polchinski, P.~Saad, S.~H. Shenker,
  D.~Stanford, A.~Streicher and M.~Tezuka, {{Black Holes and Random Matrices}},
  \href{http://dx.doi.org/10.1007/JHEP05(2017)118}{JHEP {\bf 05}, 118, 2017},
  [\href{http://arxiv.org/abs/arXiv:1611.04650}{{arXiv:1611.04650 [hep-th]}}].

\bibitem{Milekhin:2023bjv}
A.~Milekhin and J.~Xu, {{Revisiting Brownian SYK and its possible relations to
  de Sitter}},  2023,
  [\href{http://arxiv.org/abs/arXiv:2312.03623}{{arXiv:2312.03623 [hep-th]}}].

\bibitem{2008arXiv0803.3769V}
L.~L. {Vaksman}, {{Quantum Bounded Symmetric Domains}},
  \href{http://dx.doi.org/10.48550/arXiv.0803.3769}{arXiv e-prints
  arXiv:0803.3769, 2008},
  [\href{http://arxiv.org/abs/arXiv:0803.3769}{{arXiv:0803.3769 [math.QA]}}].

\bibitem{shklyarov1998function1}
D.~Shklyarov, S.~Sinel'shchikov and L.~Vaksman, {On function theory in quantum
  disc: A q-analogue of berezin transform},  1998.

\bibitem{shklyarov1999function2}
D.~Shklyarov, S.~Sinel'shchikov and L.~Vaksman, {On function theory in quantum
  disc: q-differential equations and fourier transform},  1999.

\bibitem{shklyarov1998function3}
D.~Shklyarov, S.~Sinel'shchikov and L.~Vaksman, {On function theory in quantum
  disc: Invariant kernels},  1998.

\bibitem{shklyarov1998function4}
D.~Shklyarov, S.~Sinel'shchikov and L.~Vaksman, {On function theory in quantum
  disc: Covariance},  1998.

\bibitem{shklyarov1999function5}
D.~Shklyarov, S.~Sinel'shchikov and L.~Vaksman, {On function theory in quantum
  disc: Integral representations},  1999.

\bibitem{klimyk2012quantum}
A.~Klimyk and K.~Schm{\"u}dgen, \emph{Quantum groups and their
  representations}.
\newblock Springer Science \& Business Media, 2012.

\bibitem{kassel2012quantum}
C.~Kassel, \emph{Quantum groups}, vol.~155.
\newblock Springer Science \& Business Media, 2012.

\bibitem{woronowicz1998compact}
S.~L. Woronowicz, {Compact quantum groups}, {Sym{\'e}tries quantiques (Les
  Houches, 1995) {\bf 845}, 98, 1998}.

\bibitem{manin1988quantum}
Y.~I. Manin, T.~Raedschelders and M.~Van~den Bergh, \emph{Quantum groups and
  non-commutative geometry}.
\newblock Springer, 1988.

\bibitem{faddeev1989quantum}
L.~D. Faddeev, {Quantum groups}, {Boletim da Sociedade Brasileira de
  Matem{\'a}tica-Bulletin/Brazilian Mathematical Society {\bf 20}, 47--54,
  1989}.

\bibitem{kitaev2018notes}
A.~Kitaev, {Notes on $\widetilde{\mathrm{sl}}(2,\mathbb{R})$ representations},
  2018.

\bibitem{vaksman2010quantum}
L.~L. Vaksman, \emph{Quantum Bounded symmetric domains}, vol.~238.
\newblock American Mathematical Soc., 2010.

\bibitem{Bernard:1989jq}
D.~Bernard and A.~LeClair, {{$q$ Deformation of SU(1,1) Conformal Ward
  Identities and $q$ Strings}},
  \href{http://dx.doi.org/10.1016/0370-2693(89)90953-2}{Phys. Lett. B {\bf
  227}, 417--423, 1989}.

\bibitem{ramanujan2011notebooks}
S.~Ramanujan, {Notebooks (2 volumes), tata institute of fundamental research,
  bombay, 1957}, {MR {\bf 20}, 6340, 2011}.

\bibitem{liskova1992clebsch}
N.~A. Liskova and A.~N. Kirillov, {Clebsch-gordan and racah-wigner coefficients
  for uq (su (1, 1))}, {International Journal of Modern Physics A {\bf 7},
  611--621, 1992}.

\bibitem{derkachev20143j}
S.~Derkachev and L.~Faddeev, {3j-symbol for the modular double slq (2,r )
  revisited},  in \emph{Journal of Physics: Conference Series}, vol.~532,
  p.~012005, IOP Publishing, 2014.

\bibitem{kirillov1990representations}
A.~Kirillov and N.~Y. Reshetikhin, {Representations of the algebra u (slq(2)),
  q-orthogonal polynomials and invariants of links}, {New developments in the
  theory of knots {\bf 11}, 202, 1990}.

\bibitem{jackiw1972introducing}
R.~Jackiw, {Introducing scale symmetry}, {Physics Today {\bf 25}, 23--27,
  1972}.

\bibitem{Blommaert:2023opb}
A.~Blommaert, T.~G. Mertens and S.~Yao, {{Dynamical actions and
  q-representation theory for double-scaled SYK}},  2023,
  [\href{http://arxiv.org/abs/arXiv:2306.00941}{{arXiv:2306.00941 [hep-th]}}].

\bibitem{Blommaert:2023wad}
A.~Blommaert, T.~G. Mertens and S.~Yao, {{The q-Schwarzian and Liouville
  gravity}},  2023,
  [\href{http://arxiv.org/abs/arXiv:2312.00871}{{arXiv:2312.00871 [hep-th]}}].

\bibitem{Collins_2004}
J.~Collins, A.~Perez, D.~Sudarsky, L.~Urrutia and H.~Vucetich, {Lorentz
  invariance and quantum gravity: An additional fine-tuning problem?},
  \href{http://dx.doi.org/10.1103/physrevlett.93.191301}{{\bf 93}, Physical
  Review Letters, 2004}.

\bibitem{drinfeld1986quantum}
V.~G. Drinfeld, {Quantum groups}, {Zapiski Nauchnykh Seminarov POMI {\bf 155},
  18--49, 1986}.

\end{thebibliography}\endgroup
\end{document}